\documentclass[preprint2]{aastex}

\usepackage{amssymb}
\usepackage{amsmath}
\usepackage{graphicx}

\newcommand{\p}    {\partial}
\newcommand{\f}    {\frac}
\newcommand{\ddz}  {\frac{\partial}{\partial z}}
\newcommand{\ddx}  {\frac{\partial}{\partial x}}
\newcommand{\nyin} {\nu_{in}}

\newcommand{\mbf}  {\mathbf}
\newcommand{\tA}   {\tau_{\mbox{\scriptsize{A}}}}
\newcommand{\cA}   {c_{\mbox{\scriptsize{A}}}}
\newcommand{\td}   {\tau_d}
\newcommand{\tO}   {\tau_{\Omega}}
\newcommand{\tAD}  {\tau_{AD}}
\newcommand{\tion} {\tau_{i}}
\newcommand{\tind} {\tau_{ind}}
\newcommand{\xstar}{x_{\ast}}   
\newcommand{\xl}   {x_l}        
\newcommand{\edot} {\dot{e}_{mag}}
\newcommand{\lO}   {\lambda_{\Omega}}

\begin{document}

\title{Fast Reconnection in a Two-Stage Process}

\author{Fabian Heitsch\altaffilmark{1}}
\author{Ellen G. Zweibel\altaffilmark{1}}
\author{{\em submitted to ApJ}}
\altaffiltext{1}{JILA, University of Colorado, Campus Box 440,
                 Boulder, CO 80309-0440, USA}

\shorttitle{Fast Reconnection in a Two-Stage Process}
\shortauthors{Heitsch \& Zweibel}

\begin{abstract}
Magnetic reconnection plays an essential role in the generation and
evolution of astrophysical magnetic fields. The best tested and most robust
reconnection theory is that of Parker and Sweet. According to this theory,
the reconnection rate scales with magnetic diffusivity $\lO$ as
$\lO^{1/2}$. In the interstellar medium, the Parker-Sweet reconnection
rate is far too slow to be of interest. Thus, a mechanism for fast reconnection
seems to be required.
We have studied the magnetic merging of two oppositely directed flux systems
in weakly ionized, but highly conducting, compressible gas. In such systems,
ambipolar diffusion steepens the magnetic profile, leading to a thin current
sheet. If the ion pressure is small enough, and the recombination of ions is
fast enough, the resulting rate of magnetic merging is fast, and independent
of $\lO$. Slow recombination or sufficiently large ion pressure leads to
slower merging which scales with $\lO$ as $\lO^{1/2}$. We derive a
criterion for distinguishing these two regimes, and discuss applications to
the weakly ionized ISM and to protoplanetary accretion disks.
\end{abstract}


\clearpage

%
%
\section{Outline of the Problem}

The Ohmic diffusion time scale in the interstellar medium (ISM) is so long 
compared to the dynamical time scale that at first sight the magnetic field
would appear to be perfectly frozen to the plasma. The ratio of these two time scales,
 -- the Lundquist number $S$ -- is typically $10^{15}<S<10^{21}$, implying Ohmic diffusion
times much longer than the age of the universe.

The weakness of Ohmic diffusion causes serious problems for Galactic dynamo theory,
which requires breaking the frozen-flux condition in order to convert small scale
fields to large scale fields. More generally, it is difficult to reconcile the
frozen-flux condition with the apparent smoothness of the magnetic field in the ISM
and the apparently turbulent nature of the interstellar velocity field.

Magnetic reconnection is a ``hybrid'' process which combines resistive and dynamical
effects, and acts at a rate intermediate between them. When the reconnection rate
depends on $S$ (usually a power-law dependence), it is said to be slow.
Fast reconnection, by definition, proceeds at a rate independent of $S$. It is generally 
accepted that in order for reconnection to cause a substantial breakdown of flux freezing
in the ISM, it must be fast. 

The reconnection theories of \citet{PAR1957} and \citet{SWE1958} have proven durable, and represent
in some sense the ``default'' models, to which later work is usually compared.
In the Parker-Sweet models, two opposing 
magnetic field systems are pressed together, forming a current sheet at the
magnetic null plane at which the field undergoes steady state, resistive
annihilation. According to this model,
the reconnection rate scales as $S^{1/2}$.

\citet{PET1964} argued for changing the geometrical setup from reconnection in a current sheet
to an X-point geometry.
In this model, the reconnection rate is proportional to $(\ln S)^{-1}$. 
However, both numerical
\citep{COW1975,BIS1986,SCH1989,BIS1994,UGA1995} and analytical studies 
\citep{KUL1998,KUL2001} cast doubt on whether Petschek's model is realizable, unless
special conditions are met \citep{FOR2001}.
Another solution based on flow geometry was proposed by \citet{LAV1999}, who argued that
fast reconnection occurs in the presence of turbulence at the resistive length scale;
see also \citet{CHZ1989}.

The solution to the fast reconnection problem could lie in physical processes outside the scope
of single-fluid magnetohydrodynamics. The high electric current density expected in the 
reconnection layer may lead to instabilities which could generate small scale electromagnetic or
electrostatic turbulence. Scattering of electrons by such turbulence provides
so-called anomalous resistivity
(see \citet{TRE2001} and references therein) and increases
 the reconnection rate.
If the current layers are sufficiently thin, electrons and ions may decouple \citep{MDD1994}.
As \citet{BSD1995} argue, this can lead to reconnection at a rate independent of the resistivity. 

\citet{DOK1995} showed that the reconnection layer can be made thin if the gas
pressure is reduced by progressive cooling. 
They found that a resistive region will
form within a cooling time multiplied by $\ln S$, and that the magnetic field is dissipated
within a few subsequent cooling times. However, as they argue, the temperature must be
reduced by a factor of $S^{1/4}$ for this to happen. Bearing in mind the value of $S$ in the
ISM, the study demonstrates very clearly {\em how} large $S$ actually is.

In this paper, we consider ambipolar diffusion \citep{MES1956} as a mechanism to reduce the
reconnection layer width. \citet{ZWE1989} investigated the effect of partial
ionization on tearing instabilities
in a sheared magnetic field. The reconnection rate is increased by a factor 
of $(\rho/\rho_i)^{1/5}$, but the
$S^{3/5}$-dependence found in the theory of \citet{FKR1963} is unchanged. 
\citet{VIL1999} considered Parker-Sweet reconnection in weakly ionized gas. They argued that
although the rate remains proportional to $S^{1/2}$, it can be enhanced by several orders of
magnitude due to recombination of ions in the resistive layer.
\citet{BRZ1994} showed that ambipolar diffusion steepens the magnetic profile in the vicinity of
a magnetic null layer, and suggested that the resulting high current density could set the stage
for magnetic reconnection. \citet{BRZ1995} -- hereafter BZ95 -- found that although ion pressure
tends to broaden the resistive layer and reduce the reconnection rate, a fast reconnection
regime exists that is accessible under ISM conditions. 

This paper is in some respects a followup of BZ95. We consider the steady state properties of a layer
of weakly ionized gas around a field reversal, 
with the magnetic field fixed at the outer boundaries. 
We show by analytical arguments and a set of numerical models, that magnetic merging can be 
either fast or slow. The merging rate depends on a single parameter combining the ratio $\beta$ of 
ion pressure to magnetic pressure, the Ohmic diffusivity $\lO$, the ambipolar diffusivity 
$\lambda_{AD}$, the recombination rate of ions,
 and the global length scale $L$ in the problem. Our result suggests
that fast merging of magnetic neutral sheets is possible in the weakly ionized interstellar gas.

In \S\ref{sec:descanalytic} we define the problem, quantify the relevant time scales, and
develop an analytical model for predicting the reconnection rates.
A numerical solution of the problem requires a solver which is able to
handle the disparate time scales; we describe the solver in
\S\ref{sec:descnumeric}. In \S\ref{sec:results}, the numerical results are analyzed and 
compared with the analytical theory. We discuss the implications for the ISM in
\S\ref{sec:physinterpret}. Section~\ref{sec:summary} is a summary and discussion.

%
%
\section{Analytical Description\label{sec:descanalytic}}

Before embarking on a mathematical representation of the problem, we describe
the physical setup and identify the critical parameters which we would like
to determine.

\begin{figure}[h]
  \plotone{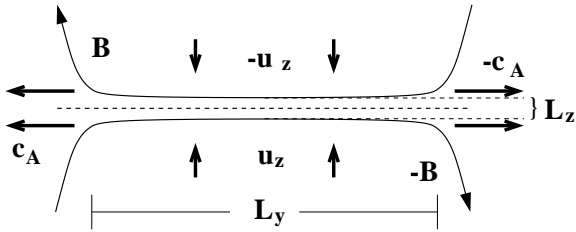}
  \caption{\label{fig:SP} Configuration for steady state reconnection according
           to Parker-Sweet theory.}
\end{figure}

We begin by summarizing reconnection theory of \citet{PAR1957}
and \citet{SWE1958}, the main features of which have been confirmed by many
subsequent calculations (e.g. \citet{BIS1993,UZK2000}). The basic
configuration is sketched in Figure~\ref{fig:SP}, 
in the coordinate system which we will
use throughout this paper.
Opposing fields $\pm\hat y B_0$ are pressed together over a length $L_y$,
leading to a current sheet of width $L_z$ centered on $z=0$. Resistive
dissipation of the current annihilates the field. 
In a steady state, magnetic flux must
be brought in at the boundaries at the rate at which it is destroyed. The
reconnection rate is parameterized by the speed $u_z$ at which the field is
convected inwards, and determining this inflow speed in terms of the
geometrical parameter $L_y$ and the intrinsic plasma
parameters is one of the central problems of
reconnection theory. 

In Parker-Sweet theory, $u_z$ is determined as follows.
The electric field $\mathbf{E}$ is assumed to be given by the standard MHD
form of Ohm's law
\begin{equation}
  \mbf{E}=-\f{\mbf{u}\times\mbf{B}}{c}+\f{\mbf{J}}{\sigma},
  \label{equ:ohmslaw}
\end{equation}
where $\sigma$ is the electrical conductivity. In the present problem, $\mbf{E}
=\hat x E$ must be constant,
because the system is in a steady state. We approximate $\mbf{E}$ by its
inductive value $-\mbf{u\times B}/c$ everywhere except in the resistive
region, where we approximate it by the expression $\mbf{J}/\sigma$. Equating
the resistive and inductive 
expressions for $\mbf{E}$, and using Amp\`{e}re's law to write $\mbf{J}$ as $\hat x
cB_0/(4\pi L_z)$, we find
\begin{equation}
  L_z=\f{\lO}{u_z},
  \label{equ:Lz1}
\end{equation}
where $\lO\equiv c^2/(4\pi\sigma)$ is the magnetic diffusivity.

 A second
relationship between $L_z$ and $u_z$ follows from conservation of mass. The
inward mass flux $\rho u_z L_y$ must be balanced by outward mass flux in the
form of thin jets within the resistive layer. Energy conservation arguments
suggest that these jets travel at the Alfv\'{e}n speed
$\cA\equiv B_0/(4\pi\rho)^{1/2}$, implying an outward
mass flux $\rho \cA L_z$. Equating the inward and outward mass fluxes yields
\begin{equation}
  L_z=L_y\f{u_z}{\cA}.
  \label{equ:Lz2}
\end{equation}
Combining equations~(\ref{equ:Lz1}) and (\ref{equ:Lz2}) yields expressions for 
$u_z$ and $L_z$
\begin{eqnarray}
  u_z=\left(\f{\lO \cA}{L_y}\right)^{1/2}=\cA S^{-1/2},
  \label{equ:uSP}\\
  L_z=\left(\f{\lO L_y}{\cA}\right)^{1/2}=L_y S^{-1/2},
  \label{equ:LSP}
\end{eqnarray}
where the Lundquist number $S$ is defined as 
\begin{equation}
  S \equiv \f{\tO}{\tA} = \f{L_y\,\cA}{\lO}.
  \label{equ:defLundquist}
\end{equation}
In the interstellar medium, $S$ is generally
enormous (quantitative estimates follow in \S\ref{subsec:dimlessformulation}), 
while the Alfv\'{e}n crossing time is comparable to the dynamical time. 
Equation~(\ref{equ:uSP}) therefore implies that reconnection is very slow.

In this paper, we express the
reconnection rate in terms of the electric field. In Parker-Sweet
reconnection, $E$ is given by
\begin{equation}
  E=\f{u_zB_0}{c}=\left(\f{\lO \cA}{L_y}\right)^{1/2}\f{B_0}{c},
  \label{equ:ESP}
\end{equation}
where we have used equation~(\ref{equ:uSP}).
The relations~(\ref{equ:uSP}) and (\ref{equ:LSP}) imply
that the Ohmic heating rate $\edot$ per unit volume is independent of $\lO$
\begin{equation}
  \edot = \f{4\pi\lO}{c^2} \mbf{J}^2 = \f{c^2}{4\pi\lO} \mbf{E}^2,
  \label{equ:magengdiss}
\end{equation}
where the second equality holds only in the resistive region. The
Poynting flux into the region, $u_zB_0^2/(4\pi)$, is proportional
to $\lO^{1/2}$.

\begin{figure}[h]
  \epsscale{0.8}
  \plotone{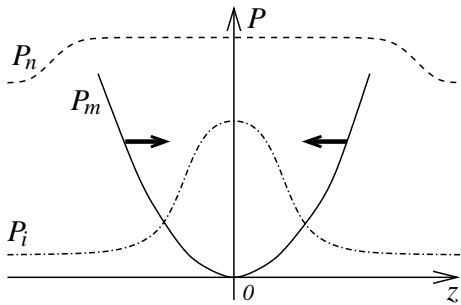}
  \caption{\label{fig:pressuresketch}Sketch of magnetic ($P_m$), ion ($P_i$) and
           neutral pressure ($P_n$) around the resistive region at $z=0$. The magnetic
           field strength $B$ would increase linearly. The neutral pressure is
           much larger than any other pressure. The magnetic pressure squeezes the
           antiparallel field lines together, only resisted by the ion pressure.}
\end{figure}

We now consider the problem in weakly ionized gas. Again, we assume a reversal
in $\hat y B(z)$ at $z=0$. When the charged and neutral matter are strongly
coupled,
regions of magnetic field reversal can achieve force balance, with neutral
pressure compensating for the deficit in magnetic pressure.
Reducing the coupling between neutrals and ions
leads to ambipolar diffusion, in which case the neutrals lag behind
the field lines. Neutral pressure support is lost, and the ion pressure
-- which is
 much smaller than the neutral pressure -- is overwhelmed by the magnetic
pressure squeezing the antiparallel field lines together.
The magnetic field gradients steepen around the magnetic null, leading
to high current densities and Ohmic dissipation. The basic situation is
sketched in Figure~\ref{fig:pressuresketch}.

An important simplification of the Parker-Sweet problem arises because the
flux of ions is not conserved. Recombination represents a sink, thus 
eliminating
the need for an outflow and making a steady state possible in one dimension.
Although the system could develop 2D or 3D structure - and, as we discuss in
\S\ref{sec:summary} this may be the only way to achieve fast reconnection under certain
conditions - we treat only the 1D case here, because it is simplest and
most tractable. In the spirit of our discussion of Parker-Sweet
reconnection, we are interested in finding the scaling of the
steady state inflow speed,
or equivalently the electric field, with the Ohmic diffusivity $\lO$.

In \S\ref{subsec:basicequations}, we write down the basic equations which govern
the system. In \S\ref{subsec:dimlessformulation}, we nondimensionalize the
problem and define and 
estimate the important basic time scales. In \S\ref{subsec:zeroiontime}
we derive the reconnection rate assuming 
ionization equilibrium is maintained at all times, and in 
\S\ref{subsec:shortiontime} we derive scaling relations for the inflow
velocity and layer thickness when the recombination rate is high but finite.

\subsection{Basic Equations\label{subsec:basicequations}}
BZ95  solved the time dependent layer problem for two fluids, one charged
and the other neutral, which are coupled by elastic collisions and by
ionization and recombination. In the parameter regime of interest, they found
that the problem is adequately described
by a reduced set of
steady state equations for the ions. We have generalized these equations to
allow for a polytropic equation of state, $P_i\propto\rho_i^{\gamma}$, instead
of the isothermal relation assumed in BZ95. 
The dynamics of the neutrals
are neglected, because the neutrals are not coupled to the magnetic
field and their pressure profile is essentially flat
(see Fig. $3$ of BZ95). The ionization fraction is sufficiently
small that the
neutrals can be regarded as an infinite reservoir which can be supplemented by
recombination with negligible effect on its overall properties. The ion flow
is assumed slow enough that ion inertia can also be neglected. As we will
see, this imposes a constraint on the width $L$ of the layers that can be
described by these reduced equations.

The magnetic induction equation, the ion momentum equation, and the ion
continuity equation are
\begin{eqnarray}
  \ddz\left(u_iB-\lO\ddz B\right)=0
  \label{equ:induction}\\
  \ddz\left(2\rho_{i0}c_{i0}^2\left(\f{\rho_i}{\rho_{i0}}\right)^\gamma
         +\f{B^2}{8\pi}\right)+\rho_i\nyin u_i=0
  \label{equ:momentum}\\
  \ddz\left(\rho_i u_i\right)-R_i\rho_{n0}+R_{rec}\rho_i^2 = 0
  \label{equ:continuity}
\end{eqnarray}
Table~\ref{tab:physcaledvars} explains the variables. The symbols $R_i$ and
$R_{rec}$ denote the ionization rate per unit time and the recombination rate
per charged particle, respectively. Equation~(\ref{equ:continuity}) is valid
for photoionization or cosmic ray ionization, and for radiative or dissociative
recombination, but not for surface recombination on grains, unless the grains
are tied to the ions rather than the neutrals.
Equation~(\ref{equ:induction}) can be integrated to yield
\begin{equation}
  u_i B-\lO\ddz B= -cE,
  \label{equ:induction-integrated}
\end{equation}
and we will always use it in this form.

Equations~(\ref{equ:momentum}), (\ref{equ:continuity}), and 
(\ref{equ:induction-integrated}) are solved on the domain $0\leq z \leq L$, 
with symmetry assumed about $z=0$. The subscripts ``0" denote conditions at $z=L$.
The boundary values at $z=0$ are $B(0)=0$, $u_i(0)=0$. 
We also impose the conditions $B(L)=B_0$ and $\rho_i(L) = \rho_{i0}$. 
We are able to impose a fourth condition in a system of three first order ordinary 
differential equations (ODEs) by treating $E$ as an eigenvalue.

For $\lO\rightarrow 0$ the order of equation~(\ref{equ:induction-integrated})
changes. Thus, we expect two physically distinct regions, namely a
resistive region around $z=0$ where $\lO\p B/\p z$ dominates, and an
inductive region for $z \gg 0$ governed by the inductive term $u_i B$, just
as occurs in the Parker-Sweet solution.

\subsection{Dimensionless Formulation and Time Scales\label{subsec:dimlessformulation}}

As the problem is largely one of time scale ratios, we
rewrite equations~(\ref{equ:momentum}) - (\ref{equ:induction-integrated})
in a dimensionless form with the relevant time scales as physical parameters.
We scale the physical variables as in Table~\ref{tab:physcaledvars}, and
introduce the time scales 
in Table~\ref{tab:timescales}. We also define the $\beta$ parameter
\begin{equation}
  \beta \equiv\f{16\pi c_{i0}^2 \rho_{i0}}{B_0^2}.
\label{equ:beta}
\end{equation}
The factor of 2 in equation~(\ref{equ:beta}) accounts for electron pressure, which
we assume to be the same as the ion pressure.

We assume ionization equilibrium at the outer boundary,
such that $\tau_{rec} = \tion$.
Rewriting the reduced equations as a set of first order ODEs for each variable 
and using equation~(\ref{equ:beta}) and Table~\ref{tab:timescales} we find
\begin{eqnarray}
  \ddx b &=& \tO\left(\f{1}{\tind} + \f{1}{\tau_A} vb\right)
  \label{equ:dimlessind}\\
  \ddx r &=& -\f{2}{\beta\gamma}r^{1-\gamma}
              \left(b \ddx b + \f{\tAD}{\tau_A}rv\right)
  \label{equ:dimlessmoment}\\
  \ddx v &=& \f{\tau_A}{\tion}\f{1}{r}(1-r^2) - \f{v}{r} \ddx r.
  \label{equ:dimlesscont}
\end{eqnarray}

Note that the time scales occur only as ratios, and that they depend on 
the length scale $L$ in different ways: $\tion$ is independent of
$L$, $\tau_A$ is linearly proportional to $L$, and $\tAD$ and $\tO$
are proportional to $L^2$. Thus, for
example, the ratio $\tO/\tau_{AD}$ is
fixed in terms of the intrinsic parameters of the medium, but the ratio of
either one to $\tion$ can be made larger or smaller by appropriate choice of
$L$.
 
Physical realism dictates certain constraints on the
time scales, which are discussed in the following subsections.

\subsubsection{The resistive time scale $\tO$}

The magnetic diffusivity $\lO$ is related to the conductivity
perpendicular to the magnetic field $\sigma_\bot$ by
\begin{equation}
  \lO= \f{c^2}{4\pi\sigma_\bot}.
  \label{equ:lambdasigma}
\end{equation}
\citet{BRA1965} gives $\sigma_\bot$ in terms of the electron collision
time $\tau_e$ as
\begin{equation}
  \sigma_\bot= \f{n_e e^2\tau_e}{m_e}.
  \label{equ:sigmaperp}
\end{equation}
Electrons collide with both charged and neutral species, at rates which we
denote by
$\tau_{ei}^{-1}$ and $\tau_{en}^{-1}$, respectively. Combining the effects of
collisions of both types leads to the composite expression
\begin{equation}
  \tau_e=\f{\tau_{ei}\tau_{en}}{\tau_{ei}+\tau_{en}}.
  \label{equ:taue}
\end{equation}
We evaluate $\tau_{ei}$ from \citet{BRA1965} with a mean ion charge $Z$
equal to unity,
and the Coulomb logarithm $\Lambda$ set equal to 20. We evaluate $\tau_{en}$
for H$_2$ as the primary neutral species, using the rates given by \citet{DRD1983}.
The result is
\begin{equation}
  \tau_e=\f{1.4\times 10^{-2}T^{3/2}}{n_e\left[1+1.2\times 10^{-11}T^2x_i^{-1}
  \right]}\,[\mbox{s}],
  \label{equ:tauenum}
\end{equation}
with the ionization fraction $x_i\equiv n_e/n_n$. Electron-neutral
collisions are represented by the second term in square
brackets in the denominator of equation~(\ref{equ:tauenum}). Such collisions are
ignorable in most of the environments considered in this
paper. However, under the conditions found in
protoplanetary disks, they dominate over electron-ion collisions.

Equations~(\ref{equ:lambdasigma}), (\ref{equ:sigmaperp}), and (\ref{equ:tauenum}) 
combine to give the magnetic diffusivity
\begin{equation}
  \lO=\f{2.0\times 10^{13}}{T^{3/2}}\left[1+1.2\times 10^{-11}T^2x_i^{-1}\right]
  \,[\mbox{cm}^2\mbox{ s}^{-1}].
  \label{equ:lambdaomega}
\end{equation}

The Ohmic diffusion time $\tO$ (see Table \ref{tab:physcaledvars}) 
is then given by 
\begin{equation}
\tO=1.54\times 10^{16}\,L_{\mbox{\small{pc}}}^2\,\f{T^{3/2}}
    {\left[1+1.2\times 10^{-11}T^2x_i^{-1}\right]}\,[\mbox{yrs}].
\label{equ:numtomega}
\end{equation}

\subsubsection{The ambipolar diffusion time scale $\tAD$}

The ambipolar diffusivity $\lambda_{AD}$
depends on the field strength $B$, the particle number densities $n_i$, $n_n$,
the molecular weights $\mu_i$ and $\mu_n$ (we assume a single species of each),
the ionization fraction $x_i=n_i/n_n$, and the rate coefficient for elastic
collisions $\langle\sigma v\rangle$.
We write
\begin{eqnarray}
  \lambda_{AD}&=&\f{(\mu_i + \mu_n)}{4\pi\langle\sigma v\rangle\mu_i\mu_n m_H x_i}
                 \left(\f{B}{n_n}\right)^2\nonumber\\
              &=&3.2\times 10^{31}\f{(\mu_i+\mu_n)}{\mu_i\mu_nx_i}
              \left(\f{B}{n_n}\right)^2\,[\mbox{cm}^2\,\mbox{s}^{-1}]\nonumber,\\
  \label{equ:deflambdaAD}
\end{eqnarray}
where we have taken $\langle\sigma v\rangle=1.5\times 10^{-9}$cm$^3$s$^{-1}$
\citep{DRD1983}.
Thus, the ambipolar diffusion time scale is (see Table \ref{tab:physcaledvars})
\begin{eqnarray}
  \tAD= 9.4\times10^{-3}\,
       L_{\mbox{\small{pc}}}^2\,\left(\f{n_n}{B}\right)^2
       \f{\mu_i\mu_n x_i}{(\mu_i+\mu_n)}\,[\mbox{yrs}].
       \label{equ:numtAD}
\end{eqnarray}

\subsubsection{The ionization time scale $\tion$\label{subsubsec:tion}}

We assume ionization equilibrium at the outer boundary so that
$\tau_{rec} = \tion$. Significant departures from ionization equilibrium are
always in the sense of a surplus of ions, so it is the recombination rates
which matter. We summarize them here.

Following \citet{DRS1987} and \citet{MCK1989}, we assume that molecular ions
are destroyed primarily by dissociative recombination, while atomic ions undergo
radiative recombination, and also recombine on the surfaces of dust grains.
Denoting the
rate coefficients for these three processes by $\alpha_{dr}$, $\alpha_{gr}$,
and $\alpha_{rr}$, respectively, we write the total recombination rate
$R_{rec}$ as
\begin{equation}
R_{rec}=(\mu_i m_H)^{-1}\left[\phi\alpha_{dr} + (1-\phi)\alpha_{rr} + \alpha_{gr}x_i^{-1}
\right],
\label{equ:alphat}
\end{equation}
where $\phi\equiv 0$ for atomic ions and $\phi\equiv 1$ for molecular ions.
The recombination time including all three processes is then given by
\begin{equation} 
\tau_{rec} = (R_{rec}\rho_i)^{-1}.
\label{equ:tion}
\end{equation}

Equation~(\ref{equ:alphat}) is valid when the grain density follows the neutral
density. As such, it is inconsistent with equation~(\ref{equ:continuity}),
because
the rate of recombination on negatively charged grains 
is then linear in the ion density rather than quadratic. We have not
included this possibility in our calculations, but we comment on it in \S 6. It
is possible that at least some of the grains follow the ions; see \citet{CM1993}
for a discussion of grain-gas coupling.

Taking expressions for the $\alpha$ coefficients from \citet{DRS1987}, we have
\begin{eqnarray}
\alpha_{rr}&=&\f{7.2\times 10^{-11}}{T^{1/2}}\label{equ:alpharr}\\
\alpha_{dr}&=&1.4\times 10^{-5}\,T^{-3/4}\\
\alpha_{gr}&=&\f{2.9\times 10^{-12}}{(\mu_i\,T)^{1/2}}\left(\f{3\mbox{\AA}}{a_{min}}\right)^{3/2}
\label{equ:alphas},
\end{eqnarray}
all in units of cm$^3$s$^{-1}$. In equations~(\ref{equ:alphas}), $a_{min}$ is the minimum 
grain size, with the size spectrum assumed to follow the MRN power law \citep{MRN1977}.
For the numbers given in Table~\ref{tab:phystimescales}, we used
 $a_{min}=3\mbox{ \AA}$.
Although this value of $a_{min}$ is uncertain, our final results are
rather insensitive to it, because of the relatively large coefficients 
of radiative and dissociative recombination. 

\subsubsection{The Alfv\'{e}n crossing time $\tA$}

The non-dimensionalization leaves us
with a ``crossing time'' in the problem. There are various ways to define
this crossing time. As it is not clear what the
velocity at the outer boundary will be (see below as well),
but values for the magnetic field
$B_0$ and the density $\rho_{i0}$ are given, we scale the velocity in terms of the
ion Alfv\'{e}n speed $c_{Ai}$ at the outer boundary $x=1$,
thus introducing the Alfv\'{e}n crossing time of Table \ref{tab:physcaledvars}.
We have
\begin{equation}
  c_{Ai}=2.2\times 10^{11}\f{B}{\left(\mu_in_nx_i\right)^{1/2}}\,[\mbox{cm}\mbox{ s}^{-1}].
  \label{equ:cAinum}
\end{equation}
The corresponding crossing time is
\begin{equation}
  \tA=0.46\,L_{\mbox{\small{pc}}}\f{\left(\mu_in_nx_i\right)^{1/2}}{B}\,[\mbox{yrs}].
  \label{equ:tA}
\end{equation}

\subsubsection{The inductive time scale $\tind$}

The inductive time $\tind$ plays a special role. Unlike the other time scales, it
is not a free parameter, but acts as an eigenvalue of the
set of ODEs~(\ref{equ:dimlessind} - \ref{equ:dimlesscont}), allowing us
to impose four instead of three boundary conditions.
We define the inductive time by the scaled electric field
\begin{equation}
  \tind\equiv{\cal E}^{-1}=\f{LB_0}{cE}.
\end{equation}

\subsubsection{The length scale $L$\label{subsubsec:lengthscale}}

We have not specified the length scale $L$ in
equations~(\ref{equ:numtomega}) and (\ref{equ:numtAD}). It is controlled
in part by the global setup of the problem, but is governed by certain constraints. 
We require $L$ to be small enough that the neutrals are decoupled from the
magnetic field, but large enough that the ion-neutral collision time 
$\nu_{in}^{-1}$ is shorter than the local ion dynamical time. 
In the subsequent discussion, we leave $L$ unspecified, in order to keep the argument
as general as possible. In \S\ref{sec:physinterpret} we 
discuss the physical systems where we believe the combination of
ambipolar diffusion and reconnection to be relevant.
Table~\ref{tab:phystimescales} contains characteristic numbers for
diffuse clouds, dense clouds, cores and protoplanetary disks.
The numbers demonstrate the disparity of time scales in the problem.

\subsection{Zero Ion Pressure Gradient\label{subsec:zeroiontime}}

If there is no ion pressure gradient, the dynamics are extremely simple, and
serve as a reference to which more general solutions can be compared. This
limit is achieved in the case $\beta\equiv 0$ and/or in the case
$\tion\equiv 0$. We refer to this solution as the pure ambipolar 
diffusion (AD) solution.

According to equation~(\ref{equ:dimlessmoment})
\begin{equation}\label{equ:ambipolv1}
  v = \f{\tau_A}{\tAD}\,b\ddx b\,\equiv\,v_{AD}.
\end{equation}
With equation~(\ref{equ:ambipolv1}), the magnetic induction 
equation~(\ref{equ:dimlessind}) can be integrated to yield
\begin{equation}\label{equ:noprs-induction}
  \f{1}{3\tAD}b^3 + \f{1}{\tO}b = \f{x}{\tind} + C.
\end{equation}
Since $b=0$ at $x=0$, the constant of integration $C$ must be zero.
Using the boundary condition $b(1)=1$ and assuming $\tO\gg\tAD$, we get
for the scaled electric field its AD-solution, namely
\begin{equation}\label{equ:noprs-efldAD}
  \frac{cE}{LB_0}=\f{1}{\tind}=\frac{1}{3\tAD}.
\end{equation}
Thus, under the conditions of pure AD, with $\lO/\lambda_{AD}\ll 1$, $E$ is 
independent of $\lO$, so reconnection is fast.

Equation~(\ref{equ:noprs-induction}) predicts that the solution has a boundary
layer near $x=0$. Far from the origin,  
\begin{eqnarray}
  b_{AD}&\approx& x^{1/3}\nonumber\\ 
  v_{AD}&\approx& - \f{\tA}{3\tau_{AD}}x^{-1/3}.
  \label{equ:ambipolb}
\end{eqnarray}
Near the origin,
\begin{eqnarray}
 b_{AD}&\approx&\f{\tO}{3\tAD}x\nonumber\\
 v_{AD}&\approx& - \frac{\tau_A}{\tAD}\left(\frac{\tO}{3\tAD}\right)^2 x.
\label{equ:resistivev}
\end{eqnarray}
Equation~(\ref{equ:ambipolb}) holds for $x> x_{AD}$, where
\begin{equation}\label{equ:ambipolzl}
  x_{AD} \equiv \left(\frac{3\tAD}{\tO}\right)^{3/2}.
\end{equation}
For $x<x_{AD}$, the solution is given by equation~(\ref{equ:resistivev}).

As equations~(\ref{equ:ambipolb}) and (\ref{equ:resistivev}) show, the flow
speed increases toward the origin, then plunges to zero. Thus, in
the linear region, the ions are
subject to tremendous compression, raising the possibility of strong
deceleration of the flow by the ion pressure gradient. In order to see how
this limits the rate of reconnection, we consider the role of ion pressure. 

\subsection{The Effect of Ion Pressure\label{subsec:shortiontime}}

As long as $\beta\ll 1$, and $\tO\ll\tAD$, ion pressure and
resistivity are important only within a boundary layer defined by
$x<\xl$. Regularity
requires $v\propto x$ as $x\rightarrow 0$, and so we introduce
a dynamical time $\td$ such that
\begin{equation}
  v=-\f{\tA}{\tau_d}x,
  \label{equ:deftd}
\end{equation}
for $x<\xl$.

Equation~(\ref{equ:deftd}) permits us to solve the continuity 
equation~(\ref{equ:dimlesscont}) for the central density $r(0)\equiv r_m$. The
result is
\begin{equation}
  r_m=\frac{\tion}{2\td}
      +\left[1+\left(\frac{\tion}{2\td}\right)^2\right]^{1/2}.
      \label{equ:defrm}
\end{equation}
We expect $r_m>1$, for large $\tO$ even $r_m \gg 1$, so that we can
write
\begin{equation}
  r_m \approx \f{\tion}{\td}
  \label{equ:defrmapprox}
\end{equation}

We now take the opportunity to correct an error in BZ95. Symmetry 
considerations require that the Taylor series for $r(x)$ contains only even
powers of $x$: $r(x) = r_m + r_2 x^2  + ...$, while the Taylor series for
$v(x)$ contains only even powers: $v(x)=-\tA x/\td - v_3 x^3-...$. BZ95
solved for $r_2$ without including $v_3$. This is incorrect, although their
conclusions as to the reconnection rate are not affected in the parameter range
they considered.

We derived equations~(\ref{equ:ambipolb}) and (\ref{equ:resistivev}) by setting
the constant of integration $C$ in equation~(\ref{equ:noprs-induction}) equal to
zero. Now, however, we are not following this solution all the way to the origin.
This allows us to introduce a parameter $\xstar$ such that
\begin{eqnarray}
  b&=&\left(\frac{x+\xstar}{1+\xstar}\right)^{1/3}\nonumber\\
  v&=&-\frac{\tA}{3\tAD\left(x+\xstar\right)^{1/3}
                          \left(1+\xstar\right)^{2/3}}.
  \label{equ:bxstar}
\end{eqnarray}
The scaled electric field is then
\begin{equation}
  {\cal E}=\frac{1}{3\,\tAD(1+\xstar)},
  \label{equ:Exstar}
\end{equation}
so $\xstar$ is effectively a parameterization of ${\cal E}$. For
$\xstar=0$ we recover the ambipolar diffusion solution equation~(\ref{equ:noprs-efldAD}), 
and this solution is approximately valid as long as $\xstar < 1$.

Our description of the system depends on three parameters which are so far
unknown. The first, $\xl$, is the location  of the transition between the inner, resistivity
and pressure dominated boundary layer and the outer, inductive region. The
second, $\td$, is the dynamical time in the inner layer. The third, $\xstar$,
measures the departure of the outer solution from the pure AD solution. We
can determine these three parameters, thereby patching together a global
solution, by applying three conditions which must connect the inner and outer
domains.

First, we assume that at $x=\xl$ the solution changes from
inductively dominated to resistively dominated. Equating the left hand side
of equation~(\ref{equ:dimlessind}) to the second term on the right hand side,
and using equation~(\ref{equ:deftd}), we find 
\begin{equation}
  \xl = \left(\f{\td}{\tO}\right)^{1/2}.
  \label{equ:xlfirstguess}
\end{equation}

As the second condition, we assume continuity of total (plasma plus magnetic)
pressure at $\xl$, and also that magnetic pressure dominates for $x>\xl$
while plasma pressure dominates for $x<\xl$. Our picture is that the plasma
pressure profile inside $x=\xl$ is quite flat. These assumptions together, along
with equations~(\ref{equ:dimlessmoment}) and (\ref{equ:bxstar}), lead to
\begin{equation}
  \beta\,r_m^\gamma = \left(\f{\xl+\xstar}{1+\xstar}\right)^{2/3}.
  \label{equ:pressurebalance}
\end{equation}

Finally, we postulate mass flux conservation across $\xl$. Using equations~(\ref{equ:deftd}) and 
(\ref{equ:bxstar}), we have
\begin{equation}
  \f{\xl}{\td} r_m = \f{1}{3\tAD\left(\xl+\xstar\right)^{1/3}
                               \left(1+\xstar\right)^{2/3}}.
  \label{equ:massfluxconserv}
\end{equation}

Eliminating $\td$ in equation~(\ref{equ:defrmapprox}) with equation~(\ref{equ:xlfirstguess})
yields
\begin{equation}
  r_m = \f{\tion}{\tO \xl^2}.
  \label{equ:rmelim}
\end{equation}
Equation~(\ref{equ:rmelim}) together with
equations~(\ref{equ:massfluxconserv}) and (\ref{equ:pressurebalance}) leads
to a pair of equations for $\xl$ and $\xstar$:
\begin{eqnarray}
  \xl^{2\gamma}(\xl+\xstar)^{2/3}(1+\xstar)^{-2/3}
  &=&\beta\left(\f{\tion}{\tO}\right)^\gamma
  \label{equ:pairxlxstar1}\\
  \xl^3(\xl+\xstar)^{-1/3} (1+\xstar)^{-2/3}&=&\f{3\tion\tAD}{\tO^2}
  \label{equ:pairxlxstar2}
\end{eqnarray}

It turns out that $\xl/\xstar\ll 1$ for 
all solutions of equations~(\ref{equ:pairxlxstar1}) and (\ref{equ:pairxlxstar2}). 
Therefore, we can approximate equations~(\ref{equ:pairxlxstar1}) and (\ref{equ:pairxlxstar2}) by
\begin{eqnarray}
  \xl^{2\gamma}\xstar^{2/3}(1+\xstar)^{-2/3}
  &=&\beta\left(\f{\tion}{\tO}\right)^\gamma
  \label{equ:pairxlxstar11}\\
  \xl^3\xstar^{-1/3} (1+\xstar)^{-2/3}&=&\f{3\,\tion\,\tAD}{\tO^2}.
  \label{equ:pairxlxstar22}
\end{eqnarray}
Raising equation~(\ref{equ:pairxlxstar11}) to the power $3/(2\gamma)$ and dividing
by equation~(\ref{equ:pairxlxstar22}) yields an equation for $\xstar$ alone
\begin{equation}
\xstar^{\f{3+\gamma}{3\gamma}}\left(1+\xstar\right)^{\f{2\gamma-3}{3\gamma}}=Z^{1/2},
\label{equ:xlone}
\end{equation}
where
\begin{equation}
  Z\equiv\beta^{3/\gamma}\f{\tion\tO}{9\tAD^2}.
  \label{equ:defbigZ}
\end{equation}

According to this analysis,
$Z$ is the fundamental parameter which controls the properties
of the layer. When $Z\ll 1$, $\xstar\ll 1$ as well, and equation~(\ref{equ:xlone})
has the approximate solution
\begin{equation}
\xstar\approx Z^{\f{3\gamma}{6+2\gamma}}\ll 1.
\label{equ:xstarcase1}
\end{equation}
When $Z\gg 1$, $\xstar\gg 1$, and equation~(\ref{equ:xlone}) yields
\begin{equation}
\xstar\approx Z^{1/2}\gg 1.
\label{equ:xstarcase3}
\end{equation}

\begin{figure}[h]
  \plotone{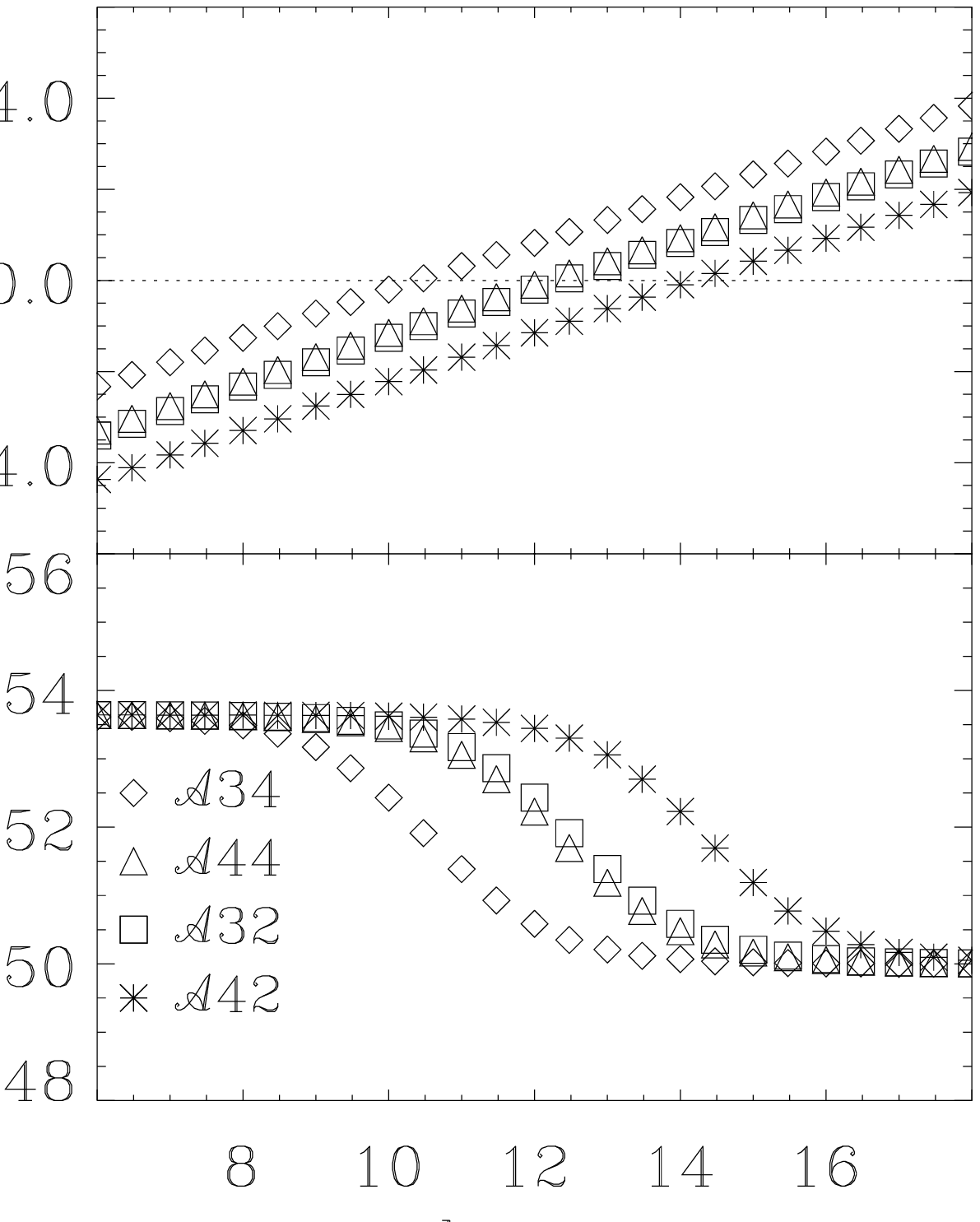}
  \caption{\label{fig:zstar-tomega}$\xstar$ against $\tO$ for the
  general solution of eqn. (\ref{equ:xlQ2}). The exponent
  change connects the regimes $\xl \ll \xstar \ll 1$
  and $\xl \ll 1 \ll \xstar$.}
\end{figure}

Equations~(\ref{equ:xstarcase1}) and (\ref{equ:xstarcase3}) agree closely when
$\gamma=5/3$, in which case the
exponent in equation~(\ref{equ:xstarcase1}) is $15/28\approx 0.54$. However, it
is useful for comparison with numerical work to find solutions of 
equations~(\ref{equ:pairxlxstar11}) and (\ref{equ:pairxlxstar22}) 
which are valid for all $\xstar
\gg \xl$. Dividing equation~(\ref{equ:pairxlxstar11}) by equation~(\ref{equ:pairxlxstar22}),
we get
\begin{equation}
\xl^{2\gamma-3}\xstar = \f{\beta}{3}\f{\tion^{\gamma-1}\tO^{2-\gamma}}{\tAD}
                        \equiv Q.
\label{equ:xlQ}
\end{equation}
We can express $\xstar$ as
\begin{equation}
  \xstar = Q\xl^{3-2\gamma}
\label{equ:xstarQ}
\end{equation}
and with equation~(\ref{equ:pairxlxstar11}) arrive at
\begin{equation}
  \xl^{3+\gamma}\left(1+Q\xl^{3-2\gamma}\right)^{-1}
  =\left(\beta\left(\f{\tion}{\tO}\right)^\gamma\right)^{3/2} Q^{-1}.
  \label{equ:xlQ2}
\end{equation}
Equation~(\ref{equ:xlQ2})
can be solved numerically for $\xl$, with $\xstar$ determined from 
equation~(\ref{equ:xstarQ}). The results are shown in Figure~\ref{fig:zstar-tomega} for parameters taken from the series of numerical models $\mathcal{A}$
listed in Table~\ref{tab:simparams}.

It is also useful to have approximate analytical expressions for
the boundary layer width $\xl$ in the
limiting cases $Z\ll 1$, $Z\gg 1$. With equations~(\ref{equ:pairxlxstar11}) and
(\ref{equ:xstarcase1}) we find
\begin{equation}
\xl=\left(9\beta\f{\tion^{\gamma+2}\tAD^2}{\tO^{\gamma+4}}\right)
        ^{1/(6+2\gamma)}
  \label{equ:xlcase1}
\end{equation}
for $Z\ll 1$. Using equations~(\ref{equ:pairxlxstar11}) and (\ref{equ:xstarcase3})
we have
\begin{equation}
\xl = \left(\beta^{1/\gamma}\f{\tion}{\tO}\right)^{1/2}
  \label{equ:xlcase3}
\end{equation}
for $Z\gg 1$.

Finally, we use equation~(\ref{equ:Exstar}) to derive the scaling of the electric
field $E$ with magnetic diffusivity $\lO$. For simplicity we assume 
$\xstar = Z^{1/2}$ for all $Z$. We find
\begin{equation}
\f{d\ln\mathcal{E}}{d\ln\lO}=\f{\sqrt{Z}}{2(1+\sqrt{Z})}.
\label{equ:escalingpred}
\end{equation}
For $Z\gg 1$, ${\cal E} \propto \lambda^{1/2}$, as in
Parker-Sweet reconnection, and for $Z\ll 1$ we expect
${\cal E}\propto\lambda^0$, i.e. ``fast'' reconnection.

%
%
\section{Numerical Method\label{sec:descnumeric}}

We seek a numerical solution of the three coupled 
ODEs~(\ref{equ:dimlessind}) - (\ref{equ:dimlesscont}). 
As we mentioned below equation~(\ref{equ:induction-integrated}), we require
that the three dependent variables $b$, $r$ and $v$ satisfy four
boundary conditions (BCs): 
\begin{eqnarray}
  b(0) = 0\label{equ:boundb0}\\
  v(0) = 0\label{equ:boundv0}\\
  b(1) = 1\label{equ:boundb1}\\
  r(1) = 1\label{equ:boundr1}
\end{eqnarray}
The first two conditions follow from symmetry arguments, and the second two
from our characterization of the equilibrium state far from the resistive
layer. Note that we do not specify $v(1)$, leaving it free to adjust to the 
varying resistivity.

Equations~(\ref{equ:dimlessind}) - (\ref{equ:dimlesscont}) together with
conditions~(\ref{equ:boundb0}) - (\ref{equ:boundr1}) constitute a two-point
boundary value problem which is also
an eigenvalue problem for $\cal{E}$.
We implement the constraint on $\cal{E}$ by introducing a fourth equation
\begin{equation}
  \ddx{\cal E} = \ddx\f{\tau_\Omega}{\tau_{ind}} \equiv 0.
\end{equation}
We solve the problem using
a combination of shooting and relaxation techniques \citep{PTV1992}

The shooting method is based on
treating the problem as an
initial-value problem with boundary conditions implemented only at $x=0$. Of
$N$ boundary conditions, let $n_1$ be imposed at $x=0$ and
$n_2=N-n_1$ BCs at $x=1$. Thus, for the initial-value problem, we have
$n_2$ freely specifiable parameters at $x=0$. Outward integration results in
values of $r$, $v$ and $b$ at $x=1$ which probably do not
fit the corresponding BCs. We zero the differences via Newton-Raphson
root finding and thus find  corrections to 
the $n_2$ free values at $x=0$. The process is repeated until the
differences at $x=1$ are below a given threshold. 

The relaxation method uses a finite-difference approximation of the ODEs on
a grid with given number of grid points. It too is based on a Newton
root-finding scheme \citep{PTV1992}.
We use the scheme in its simplest form, coupling
two neighboring grid points. Higher order coupling is possible,
but renders the method much more complex.

The relaxation method requires an initial
 guess over the full domain for all variables.
Thus, we combined both solvers, using the
shooting method to obtain an 
initial guess for the relaxation.

The length scales and time scales in this problem vary over large ranges. For
example, the ratio $\tau_\Omega/\tau_{AD}\approx 10^{10}$ for physically
realistic parameters. Therefore, our set of ODEs is very stiff. Neither the
standard shooting method nor the relaxation
method can solve the equations for
such parameters. We resolve the problem with a threefold approach.

First -- as the most obvious measure -- we use a non-uniform grid. The
standard solution would be a logarithmic
grid, however, the normalization $x(N) = 1$ comes naturally with an
exponential grid
\begin{equation}
  x(k) = e^{\delta(k-N)}.
  \label{equ:expgrid}
\end{equation}
$N$ is the number of grid points, $k$ the grid point index, and $\delta$ a
compression factor.
The boundary conditions
(eq.~\ref{equ:boundb0}--\ref{equ:boundr1}) require $b(0)=0$ and $v(0)=0$.
As we cannot reach $x=0$ with equation~(\ref{equ:expgrid}), we interpolate linearly
between a cutoff $x_{cut}$ and $x=0$. We choose $x_{cut}$ such that we do not
lose any resolution in physically interesting regions.

Second, we approach physically realistic parameters from an initial guess 
by taking small steps in parameter space. For the
$\mathcal{A}$ series of runs, we began with unrealistically small values of
$\tO$ and advanced slowly to more realistic values.

Finally, we found that at large $\tO$ or small $\beta$ the density and velocity
gradients are so steep that we can resolve them with the exponential grid
only by using an enormous number of grid points, far more than are needed over
most of the domain. Accordingly, we introduced a simple version of adaptive
mesh refinement. We refine on
the velocity difference between two adjacent grid points,
$\Delta v_k = |v_k-v_{k-1}|$. If $\Delta v_k > \Delta v_{+}$ for
$k_1 < k < k_2$, the number of grid points in this region is multiplied by
$n_{+}=8$. If $\Delta v_k < \Delta v_{-}$ for $k_1 < k < k_2$
under the condition $k_2 - k_1 > n_{-}$, the
number of grid points in this region is divided by $n_{-}=8$. 
We do not allow refinement or coarsening at the lower and upper boundaries, 
which in any case is unnecessary. Moreover, refinement is not allowed for
\begin{equation}
  \f{x_k-x_{k-1}}{x_{k-1}}\leq\epsilon_{mach},
  \label{equ:refinecondition}
\end{equation}
where $\epsilon_{mach}$ is the machine accuracy.
The $x$-grid is interpolated linearly on the new grid points. The number of
gridpoints $N$ typically varies between $10^4$ and $3\times 10^5$.

Convergence to a stable solution depends crucially on the overall error
criterion. We enforce the fractional error
$\epsilon_{conv} \equiv \Delta y/y < 10^{-4}$ for
all gridpoints, where $y$ is any of the physical variables $b$, $r$, $v$ and ${\cal E}$.
In order to ensure that we find the correct solution
for each parameter adaption iteration, the successive change of
a parameter, e.g. $\tO$, must lead to differences in the results larger than
the convergence error $\epsilon_{conv}$. Thus, $\epsilon_{conv}$
needs to be adapted with decreasing parameter adaption step size.

We have confirmed that the solution remains independent of $N$, as long as $N$ is 
large enough for the relaxation method to converge. The critical region is near
the edge of the boundary layer. Solutions for different $N$ 
are identical within the convergence error $\epsilon_{conv}$. The method develops 
oscillations in the density around $x\lesssim 1$ with $r/r_{exact} \leq \epsilon_{conv}$, 
but the overall solution is not affected.

%
%
\section{Results\label{sec:results}}

We first describe the basic properties of the solutions and
discuss the differences between simulations and analytical predictions.
Covering a range of $8$ orders of magnitude in $\tO$, we then
can infer scaling laws from the simulations and compare them to
the predictions.

In order to investigate the dependence of the scaling relations
on $\tion$ and $\beta$, we ran models ${\cal A}$ within four parameter sets,
which are listed in Table~\ref{tab:simparams}. 

\subsection{Basic Properties of the Solutions}

\begin{figure}[h]
  \plotone{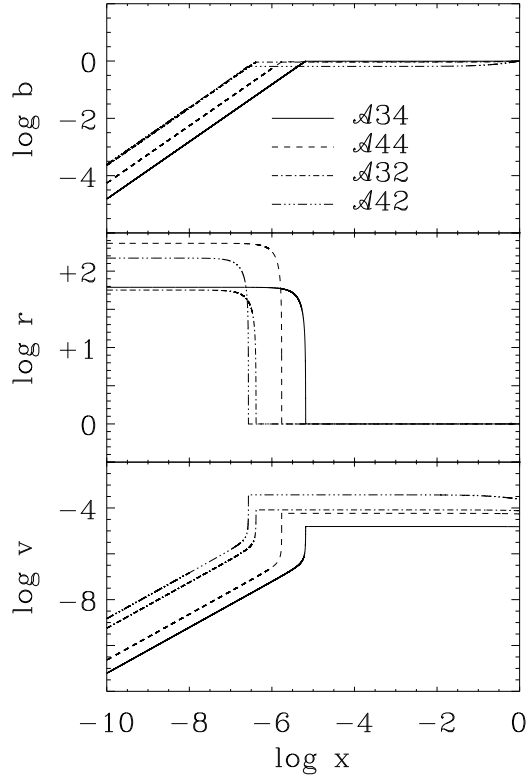}
  \caption{\label{fig:fulldomain}Magnetic field $b$, density $r$ and
          velocity $v$ at $\tO=10^{11}$. The resistive and ambipolar diffusion
          regimes are separated by a sharp transition.}
\end{figure}

\begin{figure}[h]
  \epsscale{0.75}
  \plotone{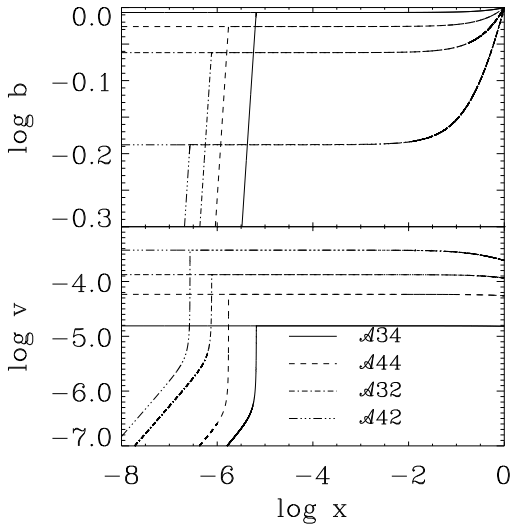}
  \caption{\label{fig:addomain}Magnetic field $b$ and
          velocity $v$ at $\tO=10^{11}$. Thick lines: numerical
          solution, thin lines: outer solution according to
          equations~(\ref{equ:ambipolb}).}
\end{figure}

Figure~\ref{fig:fulldomain} shows the full domain for all models of
type ${\cal A}$ at $\tO=10^{11}$.
Gas is moving inwards, i.e. to small $x$, at approximately constant velocity,
until it is brought nearly to a halt by the rapidly increasing plasma pressure
at $\xl$. The magnetic field and the velocity are approximately linear within the
boundary layer, as expected from equations~(\ref{equ:resistivev}), 
but essentially flat in the ambipolar or inductive regime. From equations~(\ref{equ:ambipolb}) 
we would expect $b\propto x^{1/3}$ and $v\propto x^{-1/3}$. Only for model
${\cal A}42$ do we notice a slight hint of such behavior.
However, in Figure~\ref{fig:addomain}, the overplotted outer solutions
of equations~(\ref{equ:ambipolb}) fit perfectly to
the numerical results. The fractional differences between the
numerical and analytical solutions at the transition point between the
resistive and inductive region amounts to
\begin{equation}
  \left|\f{b_{AD}-b_{sim}}{b_{AD}}\right|\approx
  \left|\f{v_{AD}-v_{sim}}{v_{AD}}\right|\approx10^{-4}=\epsilon_{conv},
\end{equation}
where the subscripts {\em AD} and {\em sim} denote the outer 
(eqs.~\ref{equ:ambipolb}) and simulation's solution. We realize that the 
solutions are so flat because, aside from a small region around $x=1$, we
have $\xstar\gg x$ for all models except ${\cal A}42$ (see
Figure~\ref{fig:zstar-tomega}).

\begin{figure}[h]
  \plotone{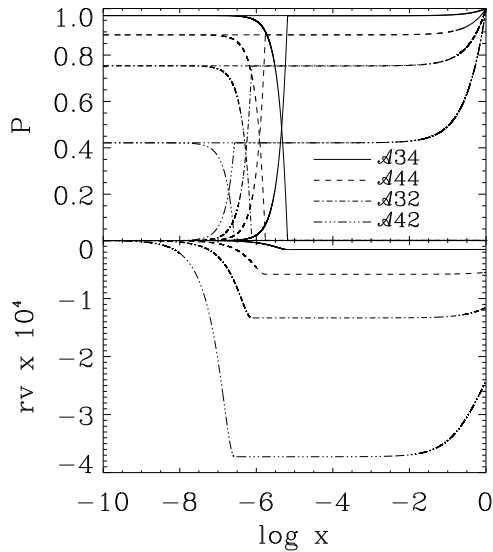}
  \caption{\label{fig:pressmasstest}{\em Upper panel}: Magnetic and ion
           pressure, and their sum for all models at $\tO=10^{11}$. 
           Magnetic and ion pressure add up to a flat curve across
           $\xl$. For $x>10^{-2}$, the magnetic pressure is
           balanced by the third term in equation~\ref{equ:momentum}.
           {\em Lower panel}: Mass flux for all models as above.
           Compared to the corresponding change in velocity (up to two orders
           of magnitude), the mass flux is nearly constant at $\xl$.}
\end{figure}

The analytical prediction of the reconnection rate (\S\ref{subsec:shortiontime})
is derived assuming pressure balance between magnetic and ion pressure
(eq.~\ref{equ:pressurebalance}) and conservation of mass flux
across the transition layer (eq.~\ref{equ:massfluxconserv}).
Figure~\ref{fig:pressmasstest} confirms these assumptions. Note that
the mass flux scale in the lower panel is linear, while the velocity scale
in Figure~\ref{fig:fulldomain} is logarithmic. While the velocity jumps
over nearly two orders of magnitude within $\Delta x\approx 10^{-7}$, the
mass flux changes at most by a factor of nearly $4$ over at least one order
of magnitude in $x$.

\begin{figure}[h]
  \plotone{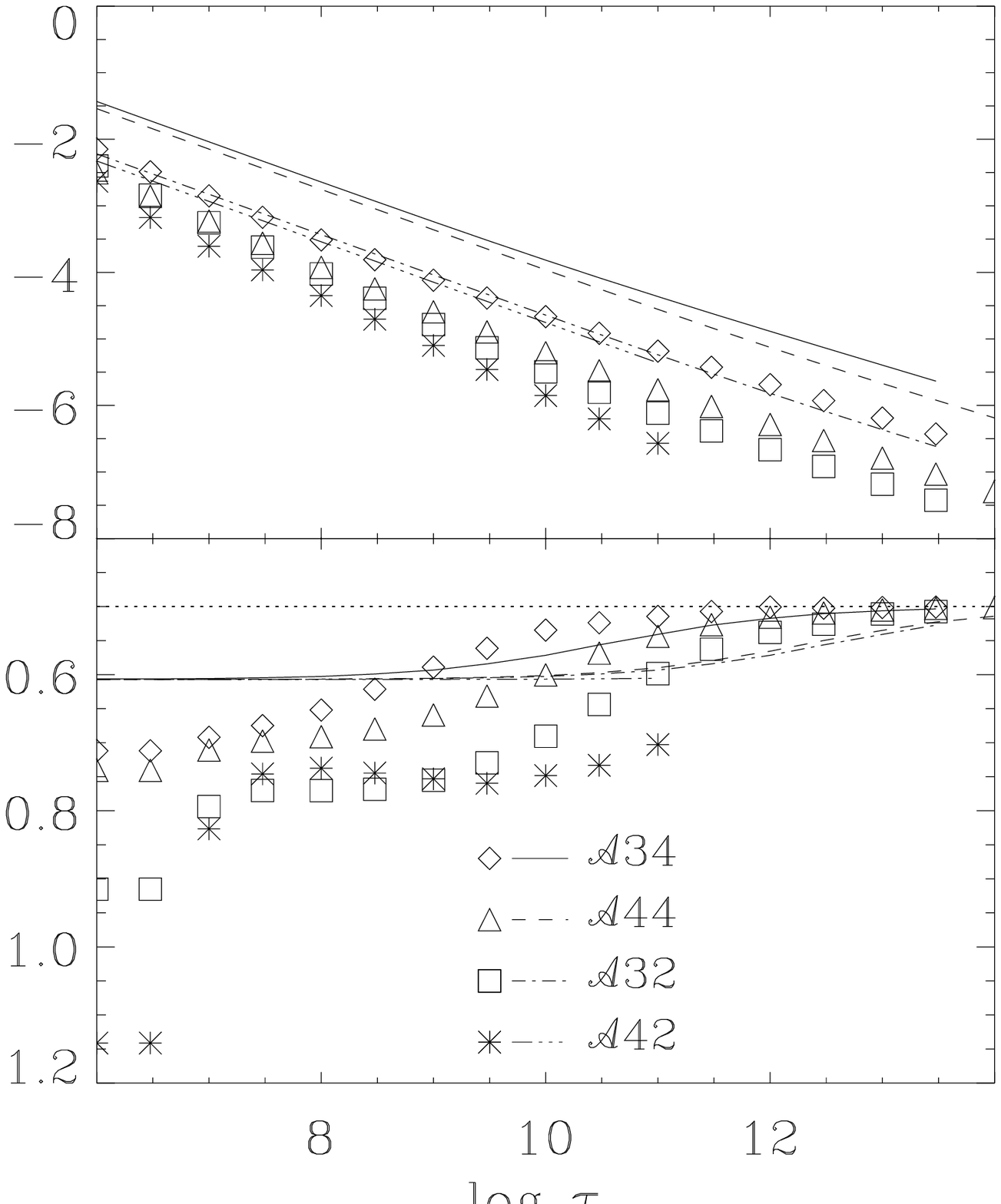}
  \caption{\label{fig:zlayer-tomega}{\em Upper panel}: Resistive scale
          $\xl$ against resistive time $\tO$, predicted (lines) according to
          equations~\ref{equ:pairxlxstar11} and \ref{equ:pairxlxstar22}, and
          determined from the simulations (symbols). {\em Lower panel}:
          Exponent $q$ of $\xl\propto\tO^q$. The dotted line
          denotes $q=-0.5$, the theoretical value for $\xstar \gg 1$.}
\end{figure}

The analysis in \S\ref{subsec:shortiontime} takes the inductive
and resistive contributions ${\cal E}_{ind}$ and
${\cal E}_{res}$ to ${\cal E}$ to be equal at $\xl$, and also uses the assumed
linearity of $v$ and $b$ with $x$. 
Figure~\ref{fig:zlayer-tomega} shows $\xl$ 
\textit{vs} $\tO$ and $d(\ln\xl)/d\ln\tO$ \textit{vs} $\tO$ for
the $\cal{A}$ simulations and for the corresponding analytical predictions.
The values of $\xl$ in the simulations can be measured accurately and
unambiguously to three significant digits from
the first or second derivatives of any of the variables $b$, $r$ and $v$. 
The predicted values of $\xl$ are derived from equation~(\ref{equ:xlQ2}).
The upper panel of Figure~\ref{fig:zlayer-tomega}
 shows a persistent offset between the simulated and predicted
values $\xl^{sim}$ and $\xl^{pre}$, in the sense that $\xl^{sim}$ is
always less than $\xl^{pre}$, sometimes by more than an order of magnitude. As
the lower panel of Figure~\ref{fig:zlayer-tomega} makes clear, the slopes of
the $\xl(\tO)$ relations differ as well, although $d(\ln\xl)/d(\ln\tO)$
computed from the simulations
converges to the predicted value of $0.5$ for large $\tO$.

In order to understand this discrepancy, we note that as long as $v$ is given
by equation~(\ref{equ:deftd}),
\begin{equation}
\xl^2=\f{{\cal E}_{ind}}{{\cal E}_{res}}\f{\td}{\tO}\f{d\ln b}{d\ln x},
\label{equ:xlgeneral}
\end{equation}
where all quantities on the right hand side are evaluated at $\xl$. 
Equation~(\ref{equ:xlgeneral}) reduces to equation~(\ref{equ:xlfirstguess}) for 
${\cal E}_{ind}/{\cal E}_{res}=1$ and $d(\ln b)/d(\ln x)=1$.

Examination of $d(\ln b)/d(\ln x)$ in our solutions shows that it drops sharply
as $x\rightarrow\xl$. With the diminished importance of the resistive term,
${\cal E}_{ind}/{\cal E}_{res}=1$ at $\xl$ actually exceeds unity for most runs.
However, $db/dx$ is steeper in the inner parts of the layer than it would
be if $d(\ln b)/d(\ln x)$ were equal to one throughout. This effectively makes
the layer thinner than equation~(\ref{equ:xlfirstguess}) predicts. 

\subsection{Reconnection Rates}

The electric field determines the magnetic reconnection rate (eq.~\ref{equ:ESP}),
and thus the magnetic energy dissipation rate per volume $\edot$ 
(eq.~\ref{equ:magengdiss}). BZ95 implicitly argued for $E\propto\lO^0$, or 
$\edot\propto\lO^{-1}$, meaning fast reconnection. They 
used $40\leq\tO/\tAD\leq4\times10^3$, resulting in a range 
$4\times10^{-3}\leq Z \leq 4\times10^1$. Thus, they were well within
the fast reconnection regime. However, they assumed isothermality, leading to 
higher ion densities in the resistive layer. 

\begin{figure}[h]
  \plotone{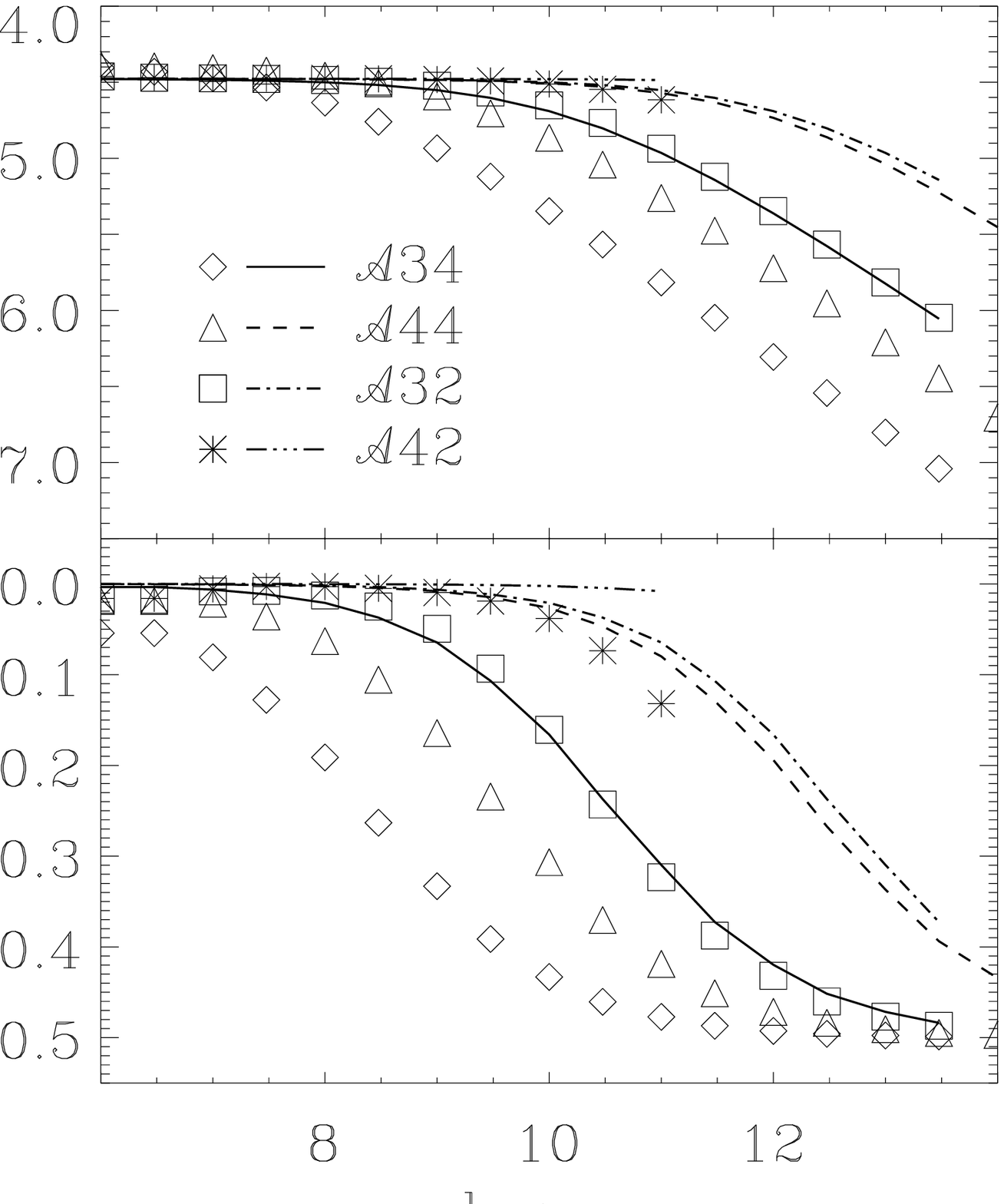}
  \caption{\label{fig:efld-lambda}Electric field ${\cal E}$ against
          resistive time $\tO$ for all models. Lines
          represent theoretical predictions, symbols stand for
          simulations. The lower panel shows
          the exponent $p$ as in ${\cal E} \propto \tO^p$.}
\end{figure}

In the previous section we saw that the scaling for $\xl$ changes at 
$\xstar \approx 1$. As Figure~\ref{fig:efld-lambda} demonstrates, 
this extends to the electric field scaling.
The $8$ orders of magnitude in $\tO$ cover the ``fast'' regime
with ${\cal E}$ independent of $\tO$ and the ``slow'' regime
with ${\cal E} \propto \tO^{-1/2}$ or equivalently 
${\cal E} \propto \lO^{1/2}$ (lower panel of
Fig.~\ref{fig:efld-lambda}). Both regimes exist. Whether they are
both equally realistic, we will discuss in \S\ref{sec:physinterpret}.

The lower panel of Figure~\ref{fig:efld-lambda} shows that the scaling 
of ${\cal E}$ with $\lO$ approaches the predicted ${\cal E}\propto\lO^{1/2}$ at large $\tO$.
However, there are two discrepancies between the simulations and the
theory. First, the value of
${\cal E}$ predicted by the analytical theory is always larger than the value
found in the simulations.
Second,  the lower panel shows that the predicted value of
$\tO$ at which the reconnection rate changes from fast to slow is
larger than the simulated value.
 
In the previous subsection, we asserted that two of the key assumptions of the
analytical theory - overall pressure balance across the layer, and conservation
of mass across the transition - are validated by the simulations. The third
assumption, equation~(\ref{equ:xlfirstguess}), is not as well supported because
of the flattening of $b$ with $x$ in the outer parts of the layer. This leads
to an overestimate of $\xl$ relative to the simulations.

\begin{figure}[h]
  \plotone{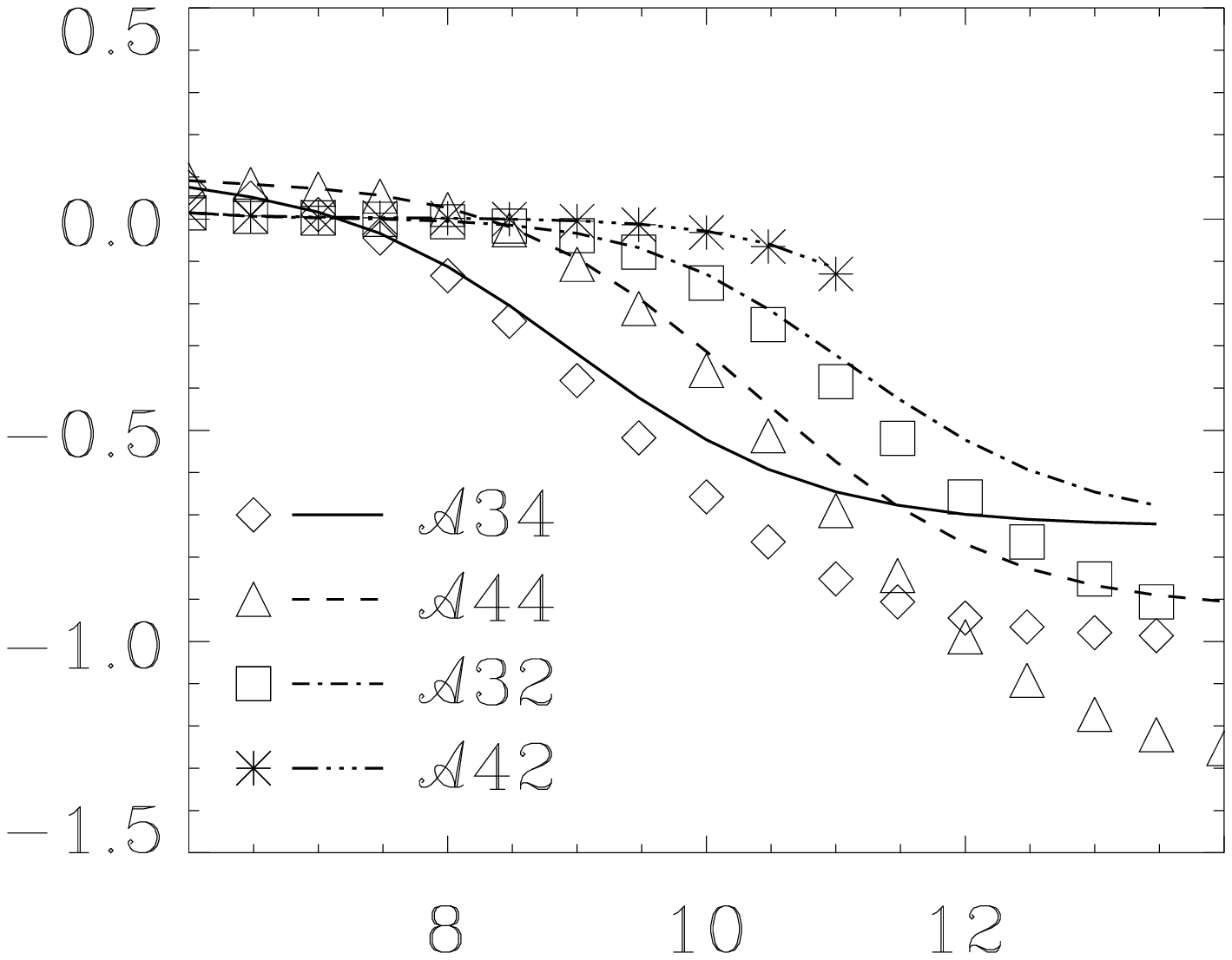}
  \caption{\label{fig:eset-lambda}Ratio of simulated over predicted electric
          field ${\cal E}$ against resistive time $\tO$ for all
          models. The agreement between theory and simulation deteriorates
          for large $\to$. Symbols denote ${\cal E}_{sim}/{\cal E}_{pre}$,
          lines stand for the same quantity with the predicted $\xl$ 
          replaced by $\xl$ from the simulations.}
\end{figure}

Much of the discrepancy between ${\cal E}_{sim}$ and ${\cal E}_{pre}$ can be 
traced to the overestimate of $\xl$ in the analytical theory 
(fig.~\ref{fig:eset-lambda}). Symbols denote ${\cal E}_{sim}/{\cal E}_{pre}$
for the ${\cal A}$ sequence models. Lines stand for the same quantity with
the predicted $\xl$ replaced by $\xl$ from the simulations. It is clear that
the agreement between theory and simulation is much better in the lower panel.

Theory and simulation agree significantly better for series
${\cal A}32$ and ${\cal A}42$ than for series ${\cal A}34$ and ${\cal A}44$
(figs.~\ref{fig:efld-lambda} and \ref{fig:eset-lambda}).
In the first two series of models, the inequality
\begin{equation}
  \tion\ll\tAD\ll\tO
  \label{equ:inequality}
\end{equation}
is well satisfied, while for the other two series it is not. 

In summary, we see a reasonable degree of consistency between the simulations
and the analytical theory. Based on their agreement, we are confident that the
theory identifies the important physical effects, while the simulations
provide numerical factors which correct the theory.

\begin{figure}[h]
  \plotone{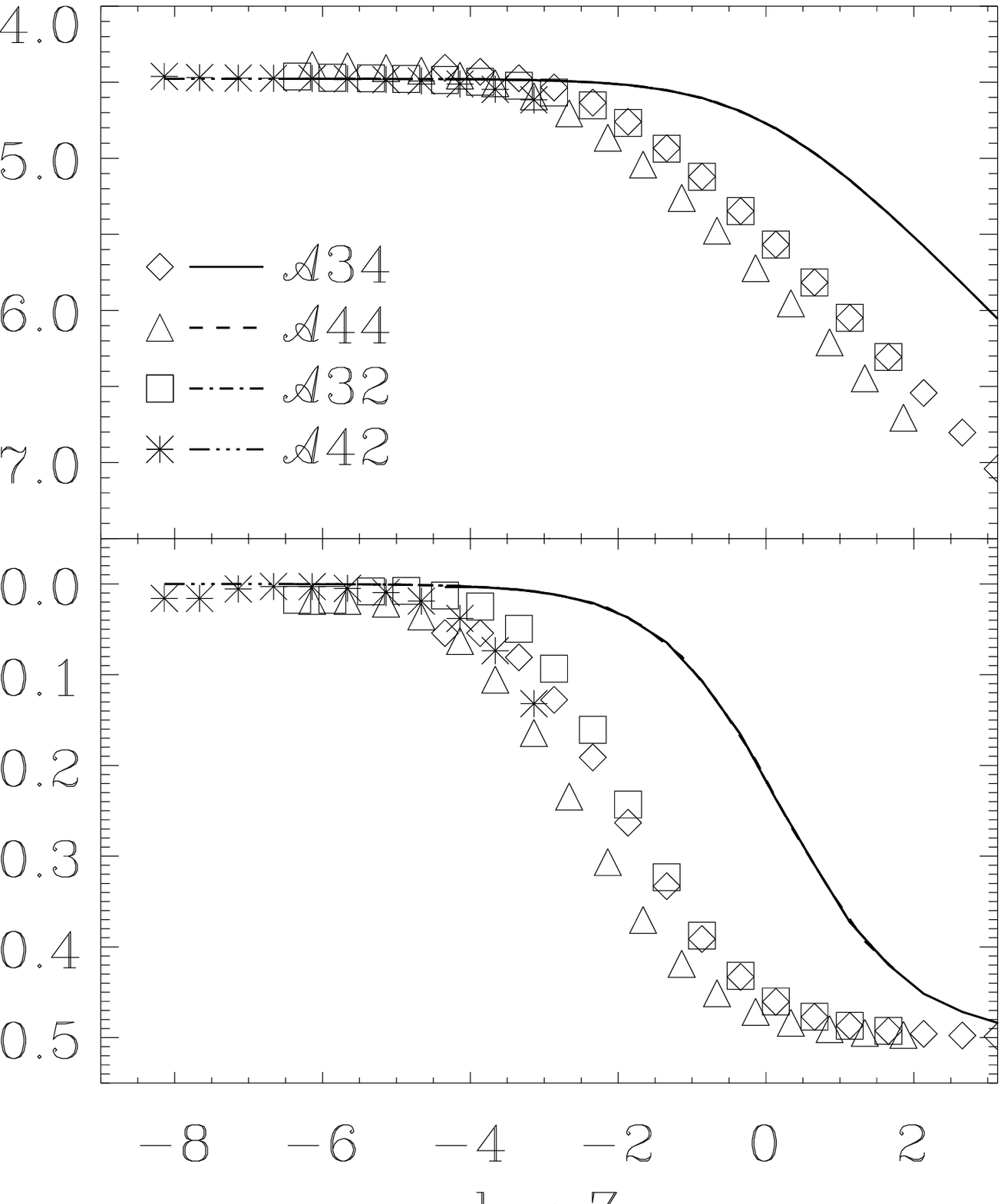}
  \caption{\label{fig:efld-bigz}Electric field ${\cal E}$ against scaling 
           parameter $Z$ as defined in equation~(\ref{equ:defbigZ}). }

\end{figure}

According to the theory, the reconnection rate is determined by the parameter
$Z$ defined in equation~(\ref{equ:defbigZ}).
Figure~\ref{fig:efld-bigz} shows ${\cal E}$ and $d\ln{\cal E}/d\ln Z$ plotted
against $Z$ for all the ${\cal A}$ series simulations (the usual symbols) and
for the theory, shown as a solid line. 
The first observation is that $Z$ is
indeed a good parameter for ordering the results of the simulations. While
the simulations do not fall onto a single curve because of the effects
discussed above (e.g. eq.~\ref{equ:inequality}), they lie quite close together.
The analytical predictions for different models are of course indiscernible.
We see also that reconnection slows down at a smaller value of $Z$ than 
predicted by the theory. Empirically, we suggest $Z\approx 0.01$ rather than
$Z=1$ as the dividing line. We identified the discrepancies between predictions 
and simulations as being caused by the over-idealized linearity of $b$ and $v$ 
for $x<\xl$. Thus, we will use the analytical predictions for the subsequent 
discussion, but correct them by replacing $Z$ by $100\,Z$.

%
%
\section{Physical Interpretation\label{sec:physinterpret}}

We now apply our results to weakly ionized astrophysical
environments. We begin by discussing the choice of length scale $L$. Then we
derive the reconnection rate as a function of $L$ for some astrophysical systems
of interest. We estimate the rate of Ohmic heating and show that it is extremely
high within the resistive layers. Finally, we discuss the implications for the
ISM and for protoplanetary disks.
\subsection{The Choice of $L$\label{subsubsec:choiceofl}}

One crucial parameter has yet to be defined: the length scale $L$ of the domain
(see \S\ref{subsubsec:lengthscale}). The scaling parameter $Z$ -- and thus
the reconnection rate -- depends on $L$ as $L^{-2}$. Thus, we predict faster
reconnection in smaller systems, which is intuitively plausible.

In order for the set of equations~(\ref{equ:induction}) - (\ref{equ:continuity}) 
to be valid, the ions and neutrals must move independently of one another.
To estimate the length scale at which this occurs, we consider
small perturbations of the two fluid system having wave number $k$ 
perpendicular to the magnetic field. In the limit $k\rightarrow 0$, the ions
and neutrals are almost perfectly coupled. As $k$ increases, the coupling
decreases, until $\mid(v_i-v_n)/v_i\mid\sim 1$ at a wavenumber $k_D$ which is
given by 
\begin{equation}
k_D=\f{2\nu_{ni}}{c_{An}}\left(1+\f{c_{sn}^2}{c_{An}^2}\right)^{1/2}.
\label{equ:kDdef}
\end{equation}

Thus, we require $L\le 2\pi/k_D\equiv L_{AD}$.
For small $c_{sn}/c_{An}$, equation~(\ref{equ:kDdef}) reverts to the usual result
for hydromagnetic waves (e.g \citet{FZS1988}). If $c_{sn}/c_{An}$ is large, as 
we might expect in the vicinity of a neutral sheet, the critical scale is 
smaller by a factor of $c_{An}/c_{sn}$.

$L_{AD}$ can also be defined in terms of the ambipolar Reynolds 
number $R_{AD}$ \citep{ZWB1997,KHM2000}
\begin{equation}
  R_{AD} = \f{Lv}{\lambda_{AD}},
  \label{equ:defreynoldsAD}
\end{equation}
where $\lambda_{AD}$ is defined in equation~(\ref{equ:deflambdaAD}) and $v$ is the
speed of the neutrals. 
Then, $L_{AD}$ is the value of $L$ for which $R_{AD}=1$. If we replace $v$ by 
the magnetosonic speed $(c_{An}^2+ c_{sn}^2)^{1/2}$, 
then $L_{AD}\approx k_D^{-1}$. However, 
equation~(\ref{equ:defreynoldsAD}) also allows for the possibility that the system is dominated
by supermagnetosonic flows.
It is possible that $L$ could be less than $L_{AD}$ if it were defined by a
geometrical property of the system, such as the thickness of an accretion disk.
However, $L$ cannot be greater than $L_{AD}$ if the equations we solved are
to be valid.

The requirement that the system be collisional gives a lower limit for $L$, namely
\begin{equation}
  L>L_{col}\equiv c_{si}/\nu_{in},.
  \label{equ:defLcol}
\end{equation}
where $\nu_{in}\equiv \mu_n n_n\langle\sigma v\rangle/(\mu_i + \mu_n)$ is the
ion-neutral collision frequency.
\subsection{Reconnection Rates in Physical Environments}

\begin{figure}[h]
  \plotone{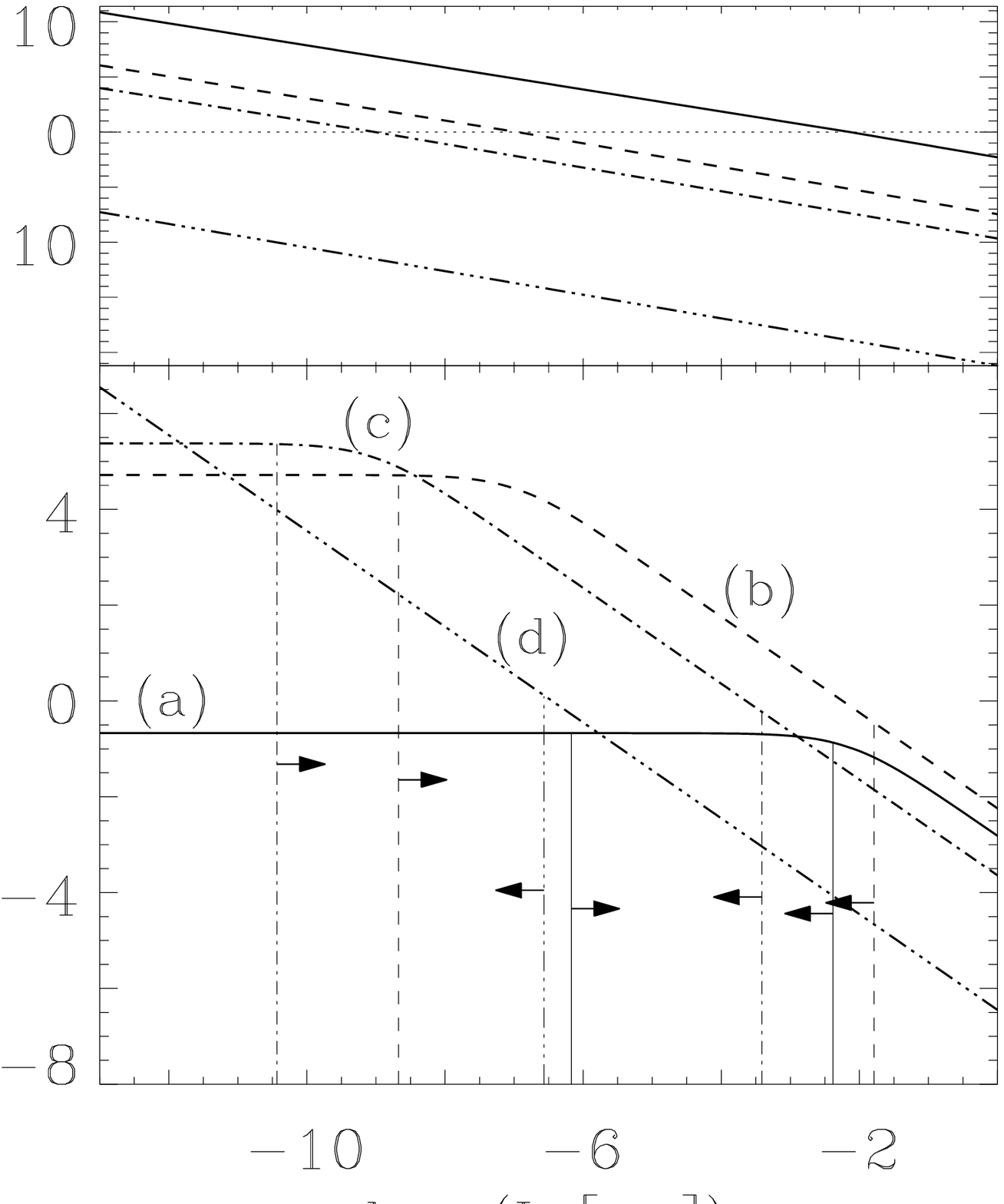}
  \caption{\label{fig:efld-length}Scaling parameter $\xstar$ and
           reconnection rate $u_i$ in units of the neutral Alfv\'{e}n speed $c_{An}$
           against domain scale for the parameter sets (a), (b), (c) and (d) as
           defined in Table~\ref{tab:phystimescales}. The vertical lines denote
           the constraints on $L$ discussed in \S\ref{subsubsec:choiceofl}, and
           right (left) pointing arrows stand for lower (upper) thresholds, according
           to equations~(\ref{equ:defreynoldsAD}) and (\ref{equ:defLcol}). We
           use the empirical formula $\xstar=10Z^{1/2}$.}
\end{figure}

Table~\ref{tab:phystimescales} lists representative values of physical
parameters in four environments: (a) diffuse clouds, (b) dense molecular clouds, (c)
molecular cores and (d) protoplanetary disks. In Figure~\ref{fig:efld-length} 
we plot the incoming ion flow speed -- the reconnection rate -- in units of 
the neutral Alfv\'{e}n speed $c_{An}$ against the scale length.

At small $L$, the curves reach a saturated, maximum value $u_{is}$. 
Using eqns.~(\ref{equ:bxstar}) and the upper panel of Figure~\ref{fig:efld-length}, 
we see that this saturated velocity is
\begin{equation}
u_{is}=\left(\f{\lO}{\tion\,\beta^{3/\gamma}}\right)^{1/2}.
\label{equ:defuis}
\end{equation}
At large $L$, eqns~(\ref{equ:bxstar}) show that $u_i$ is given approximately by 
$\lambda_{AD}/3L$. However, only the part of the domain with 
$L_{col}<L < L_{AD}$ is
accessible for any given environment. These limits are marked on each of the
curves.

Figure~\ref{fig:efld-length} shows that magnetic fields can merge quite
rapidly under realistic astrophysical conditions. Note that the 
reconnection speed $u_{is}$, which is the maximum reconnection speed as a
function of $L$ with all other parameters held fixed, actually corresponds to
\textit{slow} reconnection in the sense we originally defined it: $u_{is}
\propto \lO^{1/2}$. However, $u_{is}$ is generally much faster than the
Parker-Sweet reconnection speed $u_z$ defined in equation~(\ref{equ:uSP}). Using
equations~(\ref{equ:defLundquist}) and (\ref{equ:defuis}), we see that
\begin{equation}
\f{u_{is}}{u_z}=\f{\tau_A}{\tion\beta^{3/\gamma}},
\label{equ:speedratio}
\end{equation}
where $\tau_A$ is to be computed using the total mass density. The relative
rapidity of magnetic merging comes about because the recombination time is 
short compared to the flow time along the neutral sheet, 
and the plasma $\beta$ is very small. 

\subsection{Ohmic Heating}

If any aspect of magnetic merging of the type described here is observable, it
is likely to be related to Ohmic heating. In a steady state, the flux of 
magnetic energy into the layer is balanced by radiation, resulting in a 
radiative flux $F_r$ of
\begin{equation}
F_r=2u_i\f{B_0^2}{4\pi}.
\label{equ:radflux}
\end{equation}

\begin{figure}[h]
  \plotone{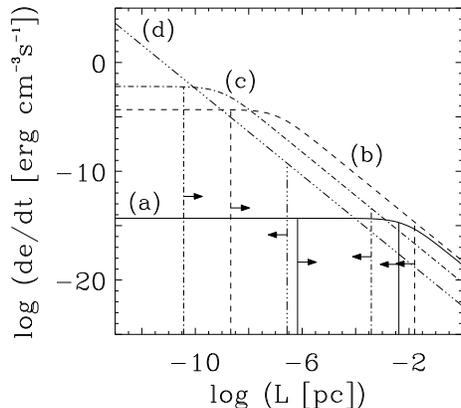}
  \caption{\label{fig:edot-length}Ohmic heating rate $\edot$ against scale length $L$
           for the parameter sets (a), (b), (c) and (d) as defined in 
           Table~\ref{tab:phystimescales}. Physically relevant regimes are denoted
           by vertical lines as in Figure~\ref{fig:efld-length}. We used the
           empirical result $\xstar=10Z^{1/2}$.} 
\end{figure}

The spectrum of the emergent radiation depends on the temperature in the layer.
While a full thermal equilibrium model is beyond the scope of this paper, it
is useful to calculate the Ohmic heating rate. At the midplane, according 
to equations~(\ref{equ:magengdiss}) and (\ref{equ:Exstar}), 
the energy dissipation rate per unit volume is
\begin{equation}
  \edot=\f{B_0^2}{4\pi\,\tion\,\beta^{3/\gamma}}\f{Z}{(1+\xstar)^2}.
  \label{equ:edot2}
\end{equation}
Recalling that $\xstar\sim Z^{1/2}$, we see that equation~(\ref{equ:edot2}) 
predicts that $\edot\propto L^{-2}$ for $\xstar\ll 1$ and
reaches a saturated value of $B_0^2/(4\pi\,\tion\,\beta^{3/\gamma})$ for $\xstar\gg 1$.
We have evaluated $\edot$ for the models in series ${\cal A}$. It
is nearly constant at
 the midplane value throughout the inner boundary layer, and plunges
to very small values outside it. Figure~\ref{fig:edot-length} shows the boundary layer
magnitudes of $\edot$ as a function of $L$ for the four environments listed in 
Table~\ref{tab:phystimescales}, with the permitted range of $L$ values indicated in each case. 
The dependence on $L$ follows the prediction of equation~(\ref{equ:edot2}).

The saturated values of $\edot$ 
in all these environments are extremely large in 
comparison to the radiative cooling rates. For example, in diffuse clouds
[case (a)], the maximum cooling rate by the 158$\mu$m fine structure
line of C~II is 
$8\times10^{-23}\mbox{ ergs cm}^{-3}\mbox{ s}^{-1}$. 
This implies that the gas is heated to very high temperatures, and
perhaps is even thermally ionized. This must slow down the 
reconnection rate at smaller values of $\tO$ than predicted by our theory. We
could very roughly account for the increase in plasma pressure by choosing 
$\gamma > 5/3$. The theory predicts that as the equation of state stiffens, the
reconnection rate slows down and the boundary layer thickens.

The high current density in the boundary
layer might destabilize high frequency waves which would provide 
anomalous resistivity through wave-particle interactions and increase the 
merging rate. 
Following the early suggestion by \citet{KRL1971},
attention focussed on the lower hybrid drift instability as the
mechanism for generating the waves. Although it now appears 
unlikely that drift waves themselves are a source of anomalous resistivity,
 but they may
play a role in the growth of Kelvin-Helmholtz instabilities, which change the
topology of the reconnection region \citep{LAB2002}. Additional
sources of small scale turbulence in null layers could also contribute to anomalous
resistivity \citep{OSH1996}.

\begin{figure}[h]
  \plotone{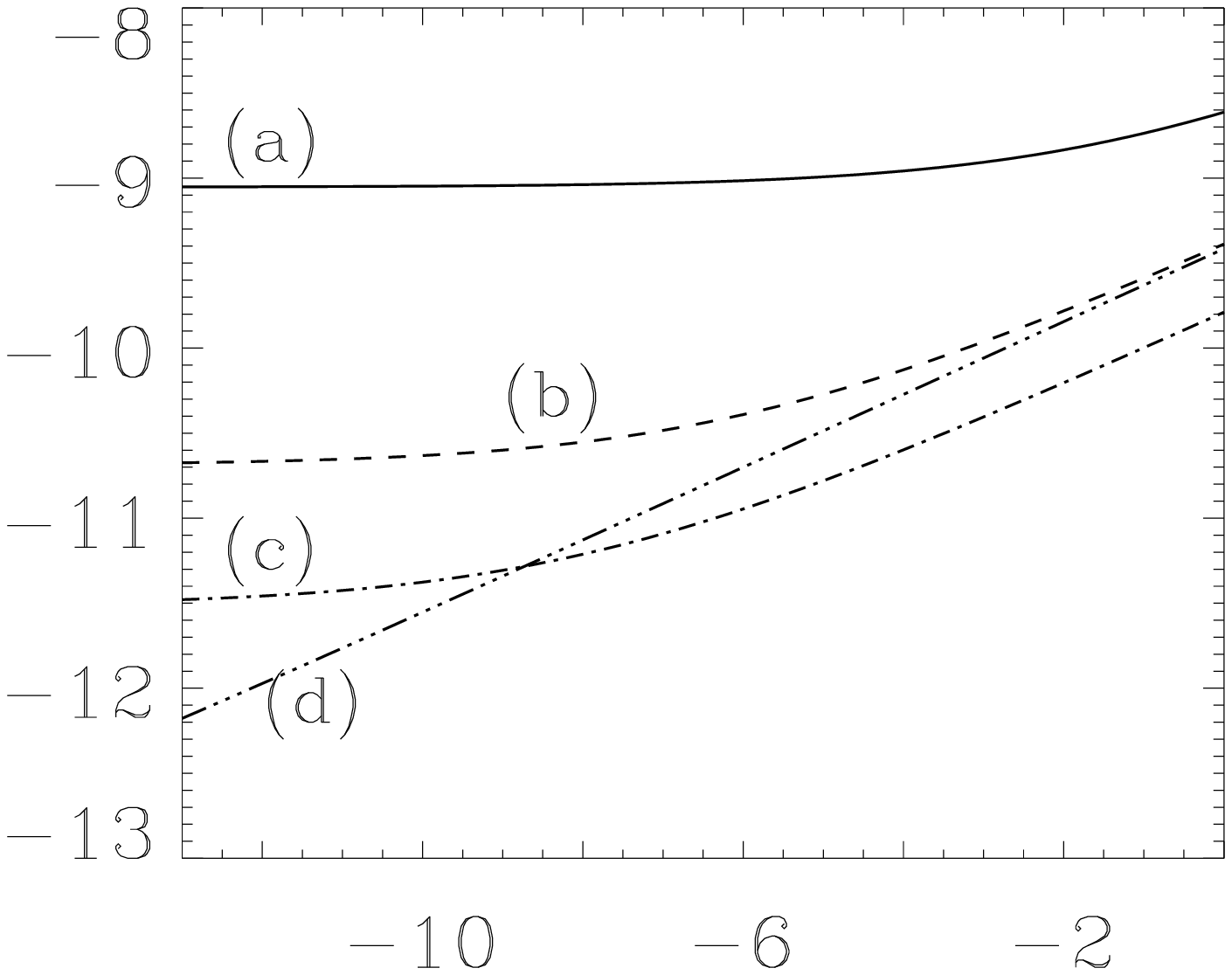}
  \caption{\label{fig:xlayer-length}Resistive layer width $L\xl$ against 
           scale length $L$ for the parameter sets (a), (b), (c) and (d) as defined in 
           Table~\ref{tab:phystimescales}. At $\xstar\approx 1$ 
           (see fig.~\ref{fig:efld-length}), the dependence of $\xl$ on $L$ 
           changes from flat to increasing. We took $\xstar=10Z^{1/2}$}
\end{figure}

Although the heating is intense, the narrowness of the layers is a
formidable barrier to their direct detectability. The analytical formulation 
predicts that the dimensional layer width, $L\xl$, scales as
$L^{3/14}$ for $Z\ll 1$ and is independent of $L$ for $Z\gg 1$; $L\xl\sim
(\beta^{1/\gamma}\,\tion\,\lO)^{1/2}$. The layer
widths are plotted in Figure~\ref{fig:xlayer-length} for the four 
environments listed in Table \ref{tab:phystimescales}.

\subsection{Reconnection in the ISM and Protoplanetary Disks}

Figure~\ref{fig:efld-length} allows us to read off the range of lengthscales
over which our theory applies, 
and rates of reconnection predicted by the theory, in a given environment.
For the diffuse clouds (a), the reconnection rate saturates at a
sub-Alfv\'{e}nic value of $u_{is}$. Within the 
permissible range of length scales
-- denoted by the vertical solid lines and inward-pointing arrows --,
$u_i$ has reached its saturated value $u_{is}$. Reconnection can occur
only if the system survives for a few tens of Alfv\'{e}n crossing times.

The situation is different for the environments (b) and (c), namely
molecular clouds and cloud cores. In both cases, the saturated reconnection
rate $u_{is}$ is super-Alfv\'enic by several orders of magnitude, and the
permissible scale length range extends well into the saturated regime. Even
in the unsaturated regime, it is possible to merge at roughly $c_{An}$ if $L$
is a few hundredths to a few thousandths of a parsec. Such scales can be imaged
directly, and are known to support density structure. Therefore, reconnection
is a viable process for the annihilation of opposing flux systems in molecular
clouds.

Protoplanetary disks are much
smaller than interstellar clouds, so curve (d) of Figure
for $L \lesssim 10^{-3}\mbox{ pc}$. The physically
permissible scale lengths range from $\approx 10^{-7}\mbox{ pc}$ down
to a few meters. The inflow velocity saturates only at the smallest
scales, but the flow velocity is super-Alfv\'{e}nic over approximately
$7$ orders of magnitude in scale length $L$. The heating provided by
fast reconnection (Fig.~\ref{fig:edot-length}) points into the direction
of models for chondrule formation as discussed by e.g. \citet{LEA1989}, 
\citet{CAM1995} and \citet{JMD2002}.

%
%
\section{Summary and Discussion\label{sec:summary}}

In this paper, we have determined the rate at which magnetic flux
can be supplied to a weakly ionized, neutral layer where it is resistively 
destroyed under steady state conditions. Through an extensive parameter 
study, covering 8 orders of magnitude in resistivity, 
we were able to find the regimes of slow and fast reconnection, 
and identify the parameter $Z$ (eq.~\ref{equ:defbigZ}) which distinguishes between them. 
Our simple analytical model and the simulations agree qualitatively.
Quantitative agreement follows directly from accounting for the over-idealized
linearity of the analytical solution. Thus, we are confident that we can apply
our model to astrophysical environments, provided that no additional effects
intervene.

Our results generalize and extend BZ95. In \S 2.4, we pointed out an error 
in the
formulation in that paper. The analysis in BZ95 predicts reconnection at a
rate independent of $\lO$ as long as $Z < \beta$, while the analysis here
predicts that the rate is independent of $\lO$ as long as $Z < 1$.

Our model is the simplest one possible. The flow is one dimensional, and the
magnetic field vanishes at the midplane. Thus, it is worth asking whether the
model can be generalized. Two dimensional flow, as in the
Parker-Sweet model of reconnection, provides another mechanism for removing
ions from the resistive layer. However, we already know that this leads to a
slow reconnection rate. Fast reconnection in our model is achieved through a
short recombination time, which removes ions more rapidly than outflow at the
Alfv\'{e}n speed. The 2D problem was considered
by \citet{VIL1999}. Their result for the reconnection rate, if Ohmic heating
is negligible, is slower than ours by a factor $\beta^{-1}$.

If we included a third component of magnetic field, which did not vanish at the
neutral line, the modifications would be more substantial. Magnetic pressure from
this third component would reduce the flow speed and prevent steepening of the
profile. In the case of a dominant third component, we should recover the
theory of tearing modes in weakly ionized gas \citep{ZWE1989}, in which 
the reconnection rate is enhanced only by a factor of $(\rho/\rho_i)^{1/5}$, 
and thus remains slow. 

Reconnection is also somewhat slowed down in situations where recombination
proceeds on grains, and the grains are tied to the neutrals. An analysis
similar to that in \S\ref{subsec:shortiontime} shows that 
$\xstar\approx\beta^{-1/2\gamma}\,Z^{1/2}$ for $\xstar\gg 1$.

Bearing these caveats in mind, fast reconnection -- i.e. a reconnection
rate independent of $\lO$ -- is possible within the model presented.
The conditions which must be met are (i) a short ionization time, guaranteeing
the fast removal of ions in the resistive layer and thus preventing
a pressure build-up, and (ii) a negligible ion pressure in comparison to
neutral and magnetic pressure. 

As we saw in Figure~(\ref{fig:efld-length}), the maximum
merging speed is actually ``slow'' in the sense that it depends on $\lO$ as
$\lO^{1/2}$. Quantitatively, however, it is quite fast: super-Alfv\'enic for
the molecular cloud and protoplanetary disk environments which we considered
(cases (b)-(d) of Table~\ref{tab:phystimescales}),
and sub-Alfv\'enic by less than two orders of magnitude for diffuse clouds
[case(a)]. We
showed that at lengthscales less than the ion-neutral decoupling length $L_{AD}$
defined by equation~(\ref{equ:defreynoldsAD}), reconnection is indeed in the
super-Alfv\'enic regime.
In protoplanetary disks (case (d)), fast reconnection might
be a good candidate to supply sufficient heat for chondrule formation.
Fundamentally, reconnection is rapid as long as the plasma $\beta$ is low and
the recombination time of ions is much less than the flow time at the Alfv\'en
speed.

An extension of the model to higher dimensions and including time dependence
seems desirable, in order to test whether the process is indeed influential
under more realistic conditions.

%
%
\acknowledgments
We would like to thank M.~Haverkorn for comments on the manuscript.
Our discussions with M.-M.~Mac Low and R.~M.~K.~Joung were very helpful. 
FH is partially supported by a Feodor-Lynen grant of the 
Alexander von Humboldt Foundation. The (U.S.) National Science
Foundation provided partial support through grants
AST-9800616 and AST-0098701.

%
%

%
%

\begin{deluxetable}{ccc}
  \tablecolumns{3}
  \tablewidth{0pt}
  \tablecaption{Physical and dimensionless variables\label{tab:physcaledvars}}
  \tablecomments{$L$ is the physical domain length, $c_{Ai}$ the
                 Alfv\'{e}n speed. Note that the scaled electric field is
                 not a dimensionless variable, but the inverse of a time scale.
                 The obvious non-dimensionalization by taking
                 $cEL/(\lO B_0)$ proves to be inconvenient because of the
                 $\lO$-dependence.}
  \tablehead{\colhead{quantity}&\colhead{physical}&\colhead{dimensionless}}
  \startdata
  length         & $z$        & $x = z/L$                     \\
  ion density    & $\rho_i$   & $r = \rho_i/\rho_{i0}$        \\
  ion velocity   & $u_i$      & $v = u_i/c_{Ai}$              \\
  magnetic field & $B$        & $b = B/B_0$                   \\
  electric field & $E$        & ${\cal E} = cE/(LB_0)$        \\
  \enddata
\end{deluxetable}

\begin{deluxetable}{ccc}
\tablecolumns{3}
\tablewidth{0pt}
\tablecaption{Time scales used in the dimensionless equations\label{tab:timescales}}
\tablecomments{Time scales used in the dimensionless equations 
 (\ref{equ:dimlessind}), (\ref{equ:dimlessmoment}, (\ref{equ:dimlessind}). The physical variables
are defined in Table~\ref{tab:physcaledvars}.}
\tablehead{\colhead{timescale}&\colhead{symbol}&\colhead{definition}}
\startdata
Ohmic & $\tO$ & $L^2/\lO$ \\
ambipolar & $\tAD$ & $L^2/\lambda_{AD}$ \\
ionization & $\tion$ & $(R_i\rho_{n0}/\rho_{i0})^{-1}$ \\
recombination & $\tau_{rec}$ & $(R_{rec}\rho_{i0})^{-1}$ \\
inductive & $\tind$ & $LB_0/cE$ \\
Alfv\'{e}n & $\tau_A$ &$ L/c_{Ai}$ \\
\enddata
\end{deluxetable}

\begin{deluxetable}{ccccc}
  \tablecolumns{3}
  \tablewidth{0pt}
  \tablecaption{Physical parameters and time scales\label{tab:phystimescales}}
  \tablecomments{Parameters for (a) diffuse clouds, (b) dense clouds, (c) cores and
                 (d) protoplanetary disks. $\gamma=c_p/c_v = 5/3$. Molecular weights $\mu$,
                 sound speed $c_s$ and Alfv\'{e}n velocity $c_{A}$ refer to the ions
                 or neutrals, depending on their index $i$ or $n$.  The ionization time
                 is given in \S\ref{subsubsec:tion}, and the plasma $\beta$
                 refers to the ions. Equations~(\ref{equ:defreynoldsAD}) and (\ref{equ:defLcol}) 
                 define $L_{AD}$ and $L_{col}$.
                 The ion gyroradius $r_{gyro}=c\sqrt{kT\mu_i m_H}/eB$.}
  \tablehead{\colhead{quantity}&\colhead{(a)}&\colhead{(b)}&\colhead{(c)}&\colhead{(d)}}
  \startdata
  $n_n\,[\mbox{cm}^{-3}]$ &$10^2$            &$10^4$            &$10^6$             &$10^{13}$\\ 
  $\mu_n$                 &$1.3$             &$2.2$             &$2.2$              &$2.2$\\
  $\mu_i$                 &$12$              &$29$              &$29$               &$40$\\
  $x_i$                   &$10^{-4}$         &$10^{-7}$         &$10^{-8}$          &$10^{-12}$\\
  $B_0\,[\mbox{G}]$       &$10^{-5}$         &$5\times 10^{-5}$ &$2\times 10^{-4}$  &$1$\\
  $T_n\,[\mbox{K}]$       &$80$              &$10$              &$30$               &$500$\\
  $c_{si}\,[\mbox{cm/s}]$ &$3.0\times10^4$   &$6.9\times10^3$   &$1.2\times10^4$    &$4.1\times10^4$\\
  $c_{Ai}\,[\mbox{cm/s}]$ &$6.3\times10^6$   &$6.4\times10^7$   &$8.1\times10^7$    &$1.1\times10^{10}$\\
  $c_{sn}\,[\mbox{cm/s}]$ &$9.2\times10^4$   &$2.5\times10^4$   &$4.3\times10^4$    &$1.8\times10^5$\\
  $c_{An}\,[\mbox{cm/s}]$ &$1.9\times10^5$   &$7.4\times10^4$   &$3.0\times10^4$    &$4.7\times10^4$\\
  $\tion\,[\mbox{yr}]$    &$3.4\times10^3$   &$7.6\times10^0$   &$2.9\times10^{-1}$ &$1.6\times10^{-7}$\\
  $\beta$                 &$4.6\times10^{-5}$&$2.3\times10^{-8}$&$4.3\times10^{-8}$ &$2.9\times10^{-11}$\\
  $r_{gyro}\,[\mbox{cm}]$ &$2.6\times10^6$   &$2.9\times10^5$   &$1.3\times10^5$    &$1.2\times10^2$\\
  $L_{AD}\,[\mbox{pc}]$   &$4.2\times10^{-3}$&$1.6\times10^{-2}$&$3.9\times10^{-4}$ &$2.7\times10^{-7}$\\
  $L_{col}\,[\mbox{pc}]$  &$6.7\times10^{-7}$&$2.1\times10^{-9}$&$3.6\times10^{-11}$&$1.2\times10^{-17}$\\
  \enddata
\end{deluxetable}

\begin{deluxetable}{ccccccc}
  \tablecolumns{3}
  \tablewidth{0pt}
  \tablecaption{Simulation parameters\label{tab:simparams}}
  \tablecomments{Time scales (in years) and $\beta$ for
                 the simulations. For all runs of type ${\cal A}$ we have
                 $10^6 \leq \tO \leq 10^{14}$, denoted by *.
                 See \S\ref{subsubsec:lengthscale} for a discussion of the physical
                 regimes covered. The first digit in the model name indicates $\beta$,
                 the second $\tion$.}
  \tablehead{\colhead{Model}&\colhead{$\beta$}&\colhead{$\tion$}
            &\colhead{$\tAD$}&\colhead{$\tA$}&\colhead{$\tO$}&\colhead{$Z/\tO$}}
  \startdata
  ${\cal A}34$  &$10^{-3}$ &$10^4$  &$10^4$  &$10$  & * &$4.4\times 10^{-11}$  \\
  ${\cal A}44$  &$10^{-4}$ &$10^4$  &$10^4$  &$10$  & * &$7.0\times 10^{-13}$ \\
  ${\cal A}32$  &$10^{-3}$ &$10^2$  &$10^4$  &$10$  & * &$4.4\times 10^{-13}$ \\
  ${\cal A}42$  &$10^{-4}$ &$10^2$  &$10^4$  &$10$  & * &$7.0\times 10^{-15}$ \\
  \enddata
\end{deluxetable}

%
%

\end{document}